\documentclass[a4paper,twoside]{article}
\newcommand{\affil}[1]{$^{\rm #1}$}
\usepackage[authoryear]{natbib}
\bibpunct{(}{)}{;}{a}{}{,}
\usepackage{graphicx}
\usepackage{subfigure}
\usepackage{ulem}
\usepackage{amsmath}
\newcommand{\grad}{\ensuremath{^{\circ}}}
\date{} 
\title{\large\bf\flushleft
Beam Patterns of the Five-hundred-metre Aperture Spherical
Telescope: Optimisation
}
\author{\parbox{\textwidth}{\flushleft
\vspace{-0.5cm}
{\it B. Dong\affil{1,2,3} and J. L. Han\affil{1}}\\
\vspace{0.4cm}
{\small \affil{1}\,National Astronomical Observatories, CAS, Jia-20 DaTun Road, Chaoyang District, Beijing 100012, People's Republic of China}\\
{\small \affil{2}\,School of Physics, University of Chinese Academy of Sciences, Beijing 100049, China}\\
{\small \affil{3}\,Corresponding author. Email: dongbin@nao.cas.cn}}}
\begin{document}
\twocolumn[
\begin{changemargin}{.8cm}{.5cm}
\begin{minipage}{.9\textwidth}
\vspace{-1cm}
\maketitle
%
%
\small{\bf Abstract:} The Five-hundred-meter Aperture Spherical
Telescope (FAST) uses adaptive spherical panels to achieve a huge
collecting area for radio waves. In this paper, we try to explore the
optimal parameters for the curvature radius of spherical panels and
the focal distance by comparison of the calculated beam patterns. We
show that to get the best beam shape and maximum gain, the optimal
curvature radius of panels is around 300~m, and a small shift in the focal
distance of a few cm is needed. The aperture efficiency can be
improved by $\sim$10$\%$ at 3~GHz by this small shift. We also
try to optimise the panel positioning for the best beam, and find
that panel shifts of a few mm can improve the beam pattern by a
similar extent. Our results indicate that accurate control of
the feed and panel positions to the mm level is very crucial
for the stability of FAST's observational performance.
\\
\medskip{\bf Keywords:} techniques: miscellaneous --- telescopes

\medskip
\medskip
\end{minipage}
\end{changemargin}
]
\small

\section{Introduction}

\begin{figure}[h]
\includegraphics[bb=13 13 710 320,width=0.475\textwidth,height=0.21\textwidth,clip]{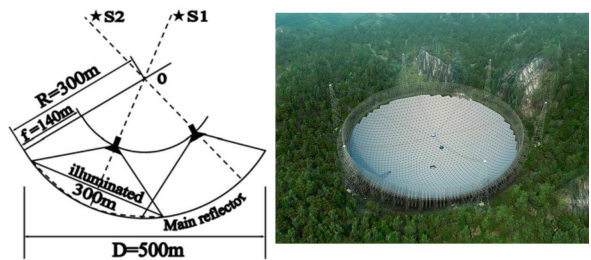}
\caption{\small FAST optical geometry in the left from \citet{nan06} and
  its 3-D model image in the right from \citet{nan11}.}
\label{fastmodel}
\end{figure}

The Five-hundred-meter Aperture Spherical Telescope (FAST) is being
constructed in a karst depression in Guizhou Province as one of the
mega science facilities for basic research in China \citep{nan11}.
The spherical panels with an overall diameter of 500~m will be used
to collect radio waves from the universe. It has about 4400 active
triangular panels with a spherical surface and a curvature radius of
318.5~m, while sitting on a cable-net that has a spherical shape and
a radius of 300~m (Figure~\ref{fastmodel}). During observations, the
illuminated panels of the main reflector will be adjusted instantaneously
to form a paraboloid with an aperture of 300~m in diameter \citep{qiu98}
operating under the real time control. The feed cabin is suspended
about 140~m above the panels, with a focal ratio of $f/D = 0.4665$.
FAST will be able to observe radio sources up to a zenith angle of
$z=40\grad$ in a frequency range of 70~MHz to 3~GHz. \citet{dh13}
recently calculated the beam patterns of FAST at 200~MHz, 1.4~GHz and
3.0~GHz for observations at zenith angles of $z=0\grad$, $27\grad$
(S1 in Fig.~\ref{fastmodel}) and $40\grad$ (S2 in Fig.~\ref{fastmodel}).

When spherical panels with the same curvature radius are used to fit a
paraboloid, the deviation of panel surface from an ideal parabolic
shape is unavoidable. Each spherical panel in FAST leads to some
axial defocusing effect, which affects the focal point and telescope
gain. The curvature radius of spherical panels and the focal distance
are therefore important parameters to optimise. We noticed that the
early design of the spherical curvature radius was 300~m \citep{qiu98},
and later it was officially designed to be 318.5~m \citep{nan11} based
on calculations of the minimum RMS deviation \citep{nan06,gj10}.
However, the illumination function of a practical feed, gaps between
panels, etc, were not taken into account, and the aperture efficiency
was therefore overestimated to be as surprisingly high as 93.3\% \citep{gj10}.

During the beam pattern calculation \citep{dh13}, we noticed that the
beam shape and telescope gain are very sensitive to the focal distance.
In that paper, we calculated the beam patterns of FAST with official
parameters in \citet{nan11}, i.e. the curvature radius of panels
$\rho_c$=318.5~m, and the focal distance $f = 300~{\rm m} \times 0.4665$ = 139.95~m.
Using a coaxial feed with an edge taper of $T_e$=$-$10.7~dB, we found
the aperture efficiency at 3 GHz as being $\sim$$57\%$, about $20\%$
lower than that of an ideal 300~m paraboloid.

In this paper, we explore the parameter space of the curvature radius
of spherical panels and the focal distance and optimise the
positions of panels for the best beam shapes and the maximum gain
by comparison of the calculated beam patterns of FAST at 3 GHz. At
this frequency the performance of FAST is more sensitive to these
parameters than at lower frequencies. In Sect.~2 we briefly describe
the FAST models and feed models used in the calculations. In Sect.~3
we show our analyses of the focal position of FAST and the curvature
radius of a paraboloid, and present the calculation results by using
the Shooting and Bouncing Ray method. Two approaches are tried
to get the best beam patterns at 3~GHz. The first one is to explore
the parameter space of the curvature radius of spherical panels and
the focal distance, and the second is to search for the best positions
of panels. Conclusions and discussions are given in Sect.~4.

\begin{figure}[!h]
\begin{center}
    \subfigure[3D model view]{
    \includegraphics[bb=2 0 345 392,width=0.180\textwidth,height=0.20\textwidth,clip]{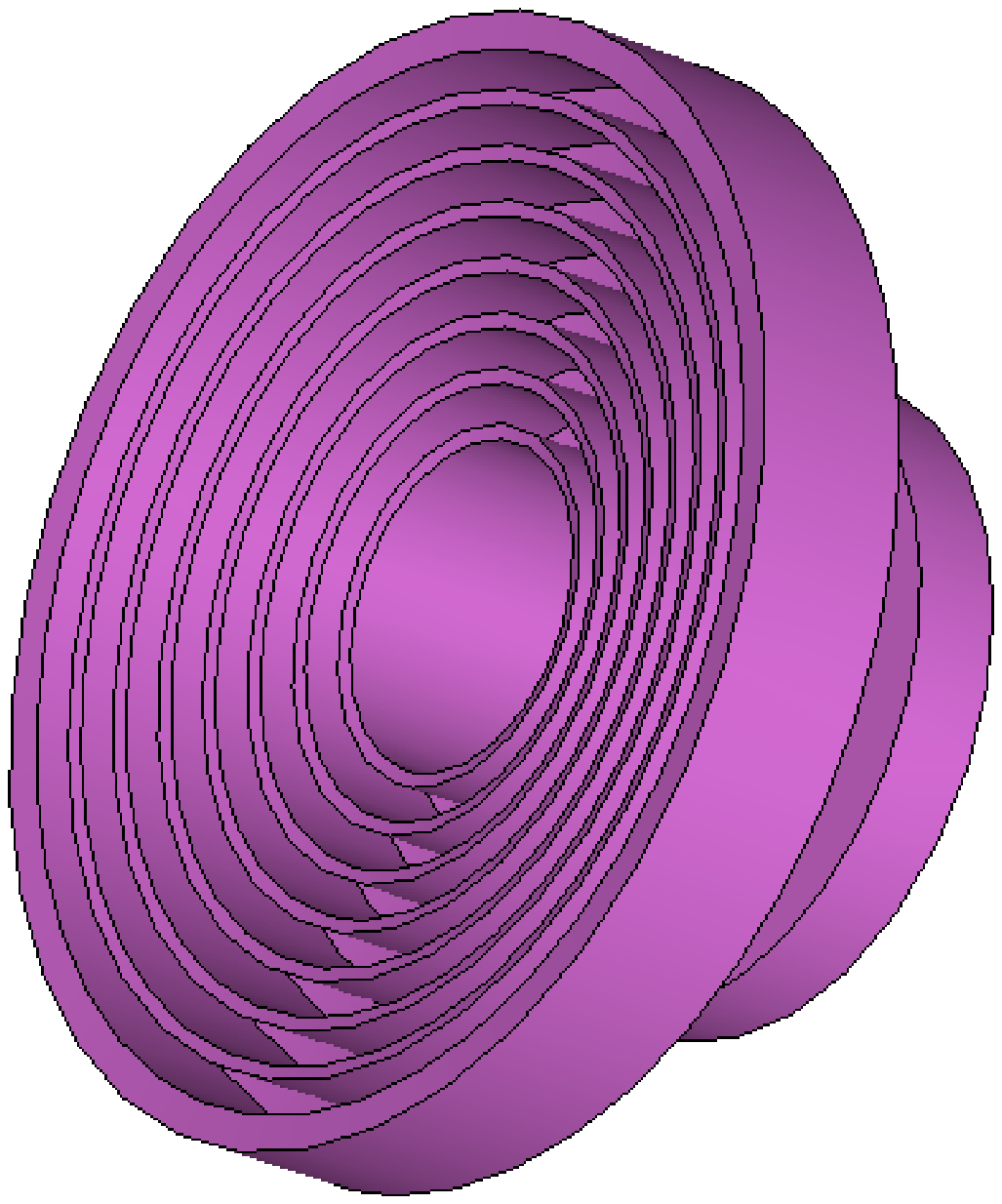}
    \label{fig2a}}
  \subfigure[Radiation pattern]{
    \includegraphics[bb=79 37 610 580,width=0.23\textwidth,height=0.23\textwidth,clip]{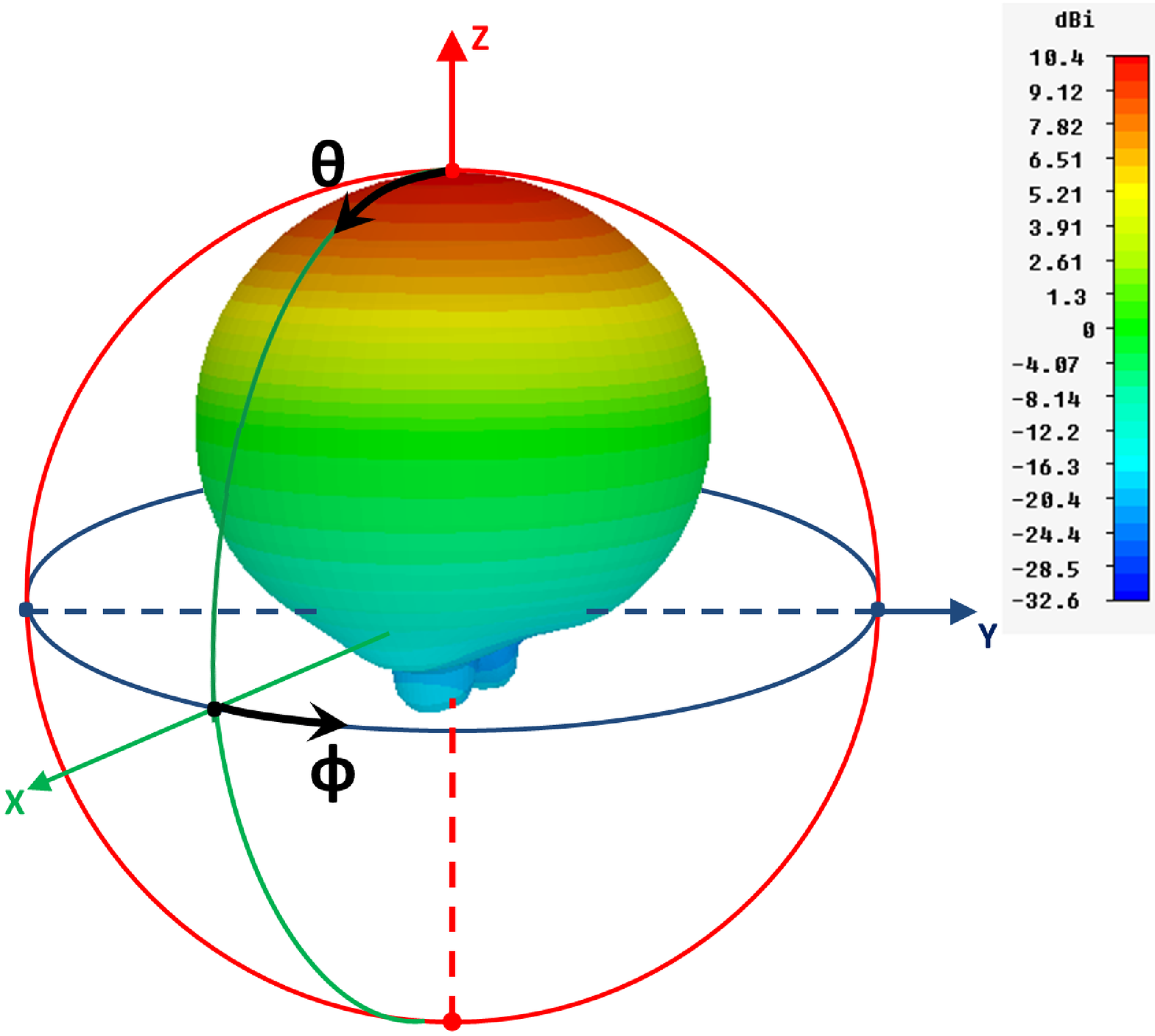}
    \label{fig2b}}
  \subfigure[Normalized patterns in the plane of $\phi$=45\grad]{
    \includegraphics[bb=2 230 574 600,width=0.43\textwidth,height=0.27\textwidth,clip]{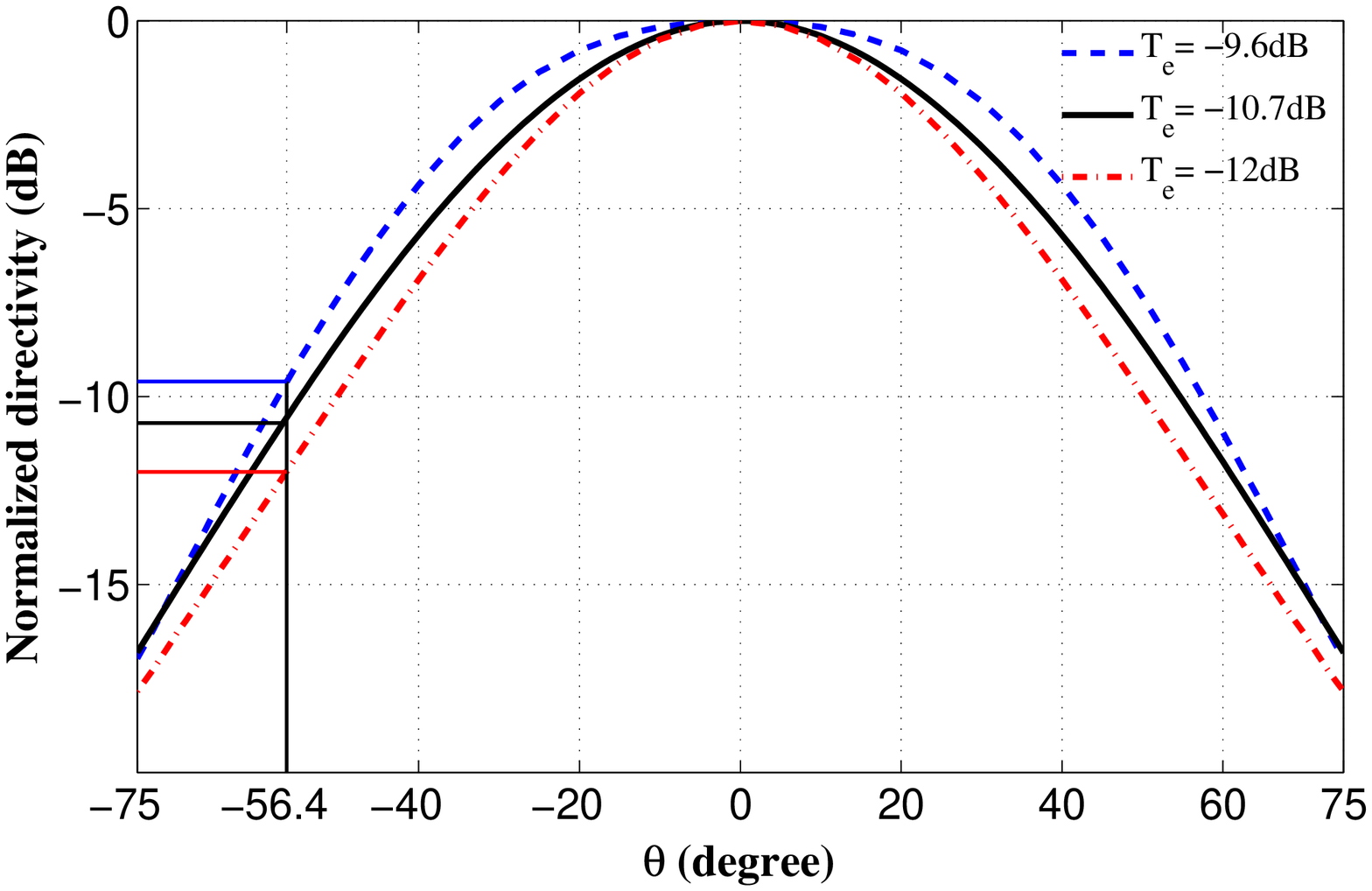}
    \label{fig2c}}
  \caption{Model of coaxial feed and radiation patterns with three edge tapers.}
  \label{fig2}
\vspace{3mm}
    \subfigure[ideal 300~m paraboloid]{
    \includegraphics[bb=90 73 442 265,width=0.215\textwidth,height=20mm]{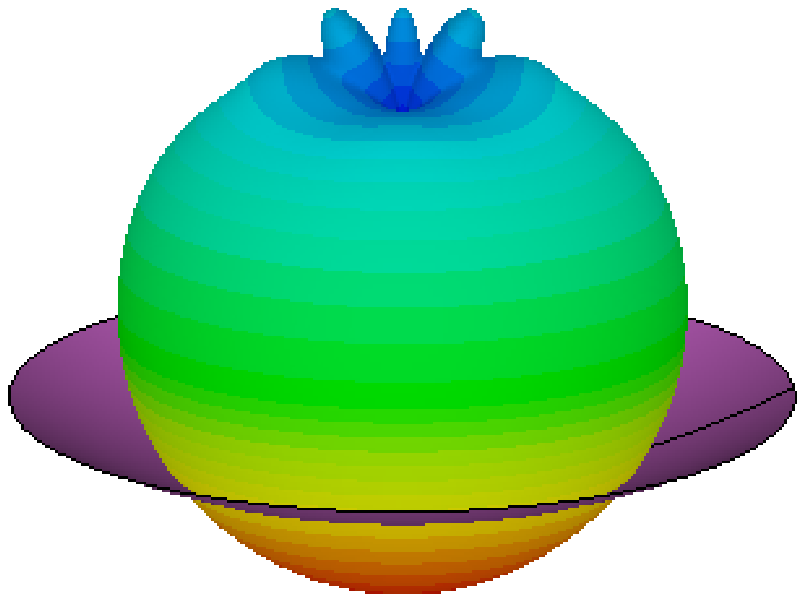}}
    \subfigure[$z$=0\grad]{
    \includegraphics[bb=15 16 527 352,width=0.215\textwidth,height=25mm]{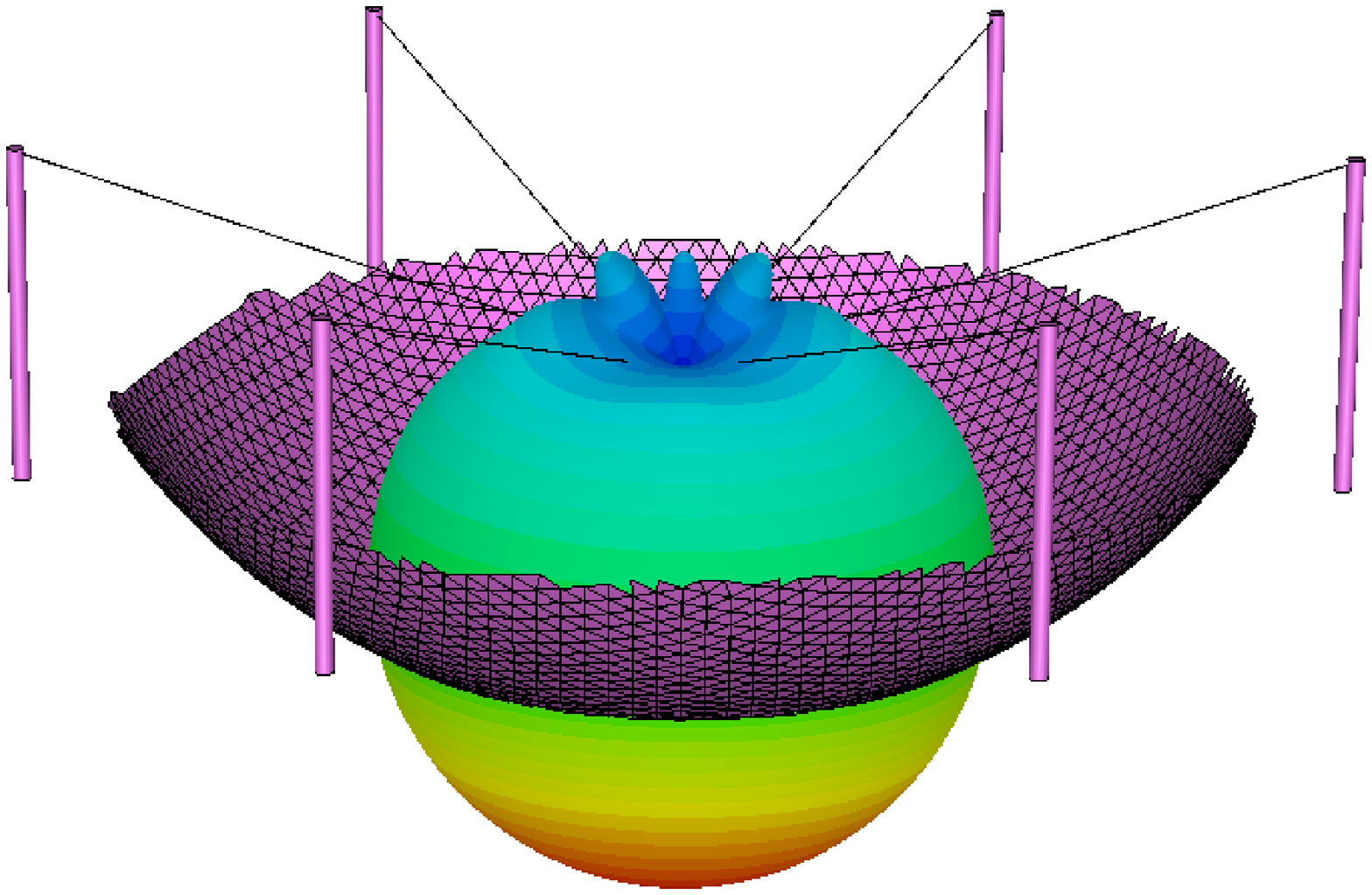}}
    \subfigure[$z$=27\grad]{
    \includegraphics[bb=14 15 553 354,width=0.215\textwidth,height=24.5mm]{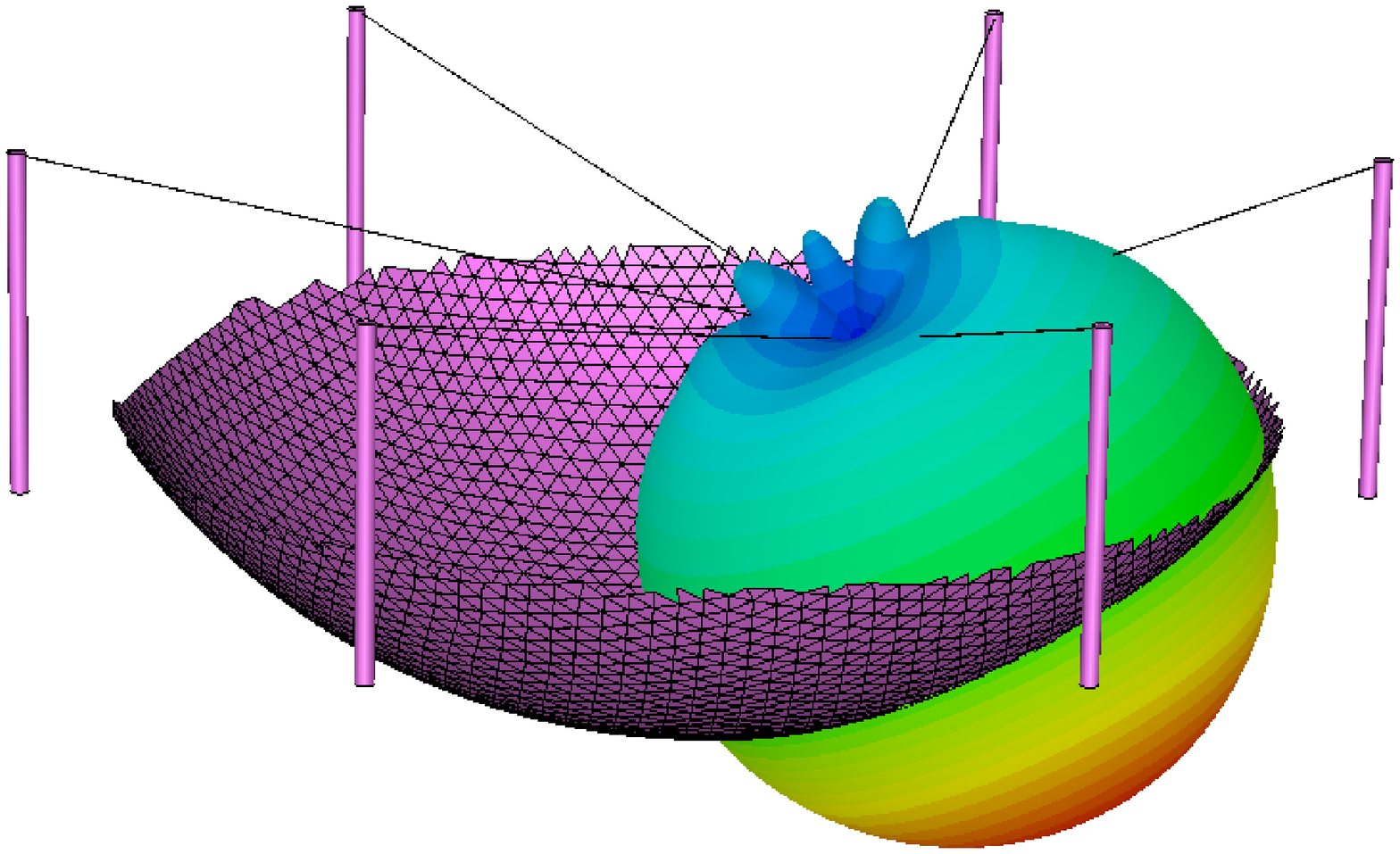}}
    \subfigure[$z$=40\grad]{
    \includegraphics[bb=0 -5 540 311,width=0.225\textwidth,height=24.5mm]{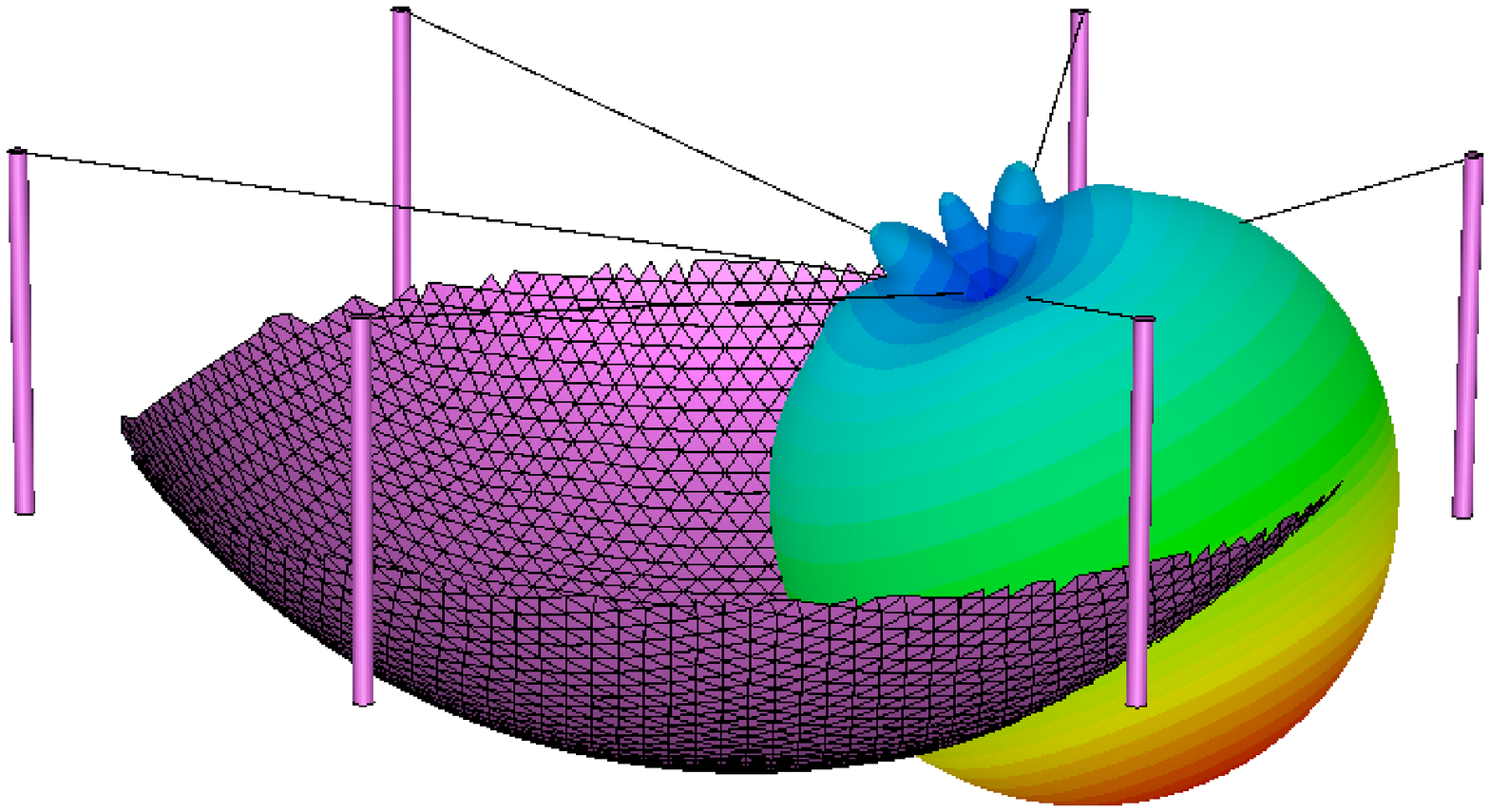}}
  \caption{Models for feed illumination of an ideal 300 m paraboloid
    and the FAST for observations at $z=0\grad$, $z=27\grad$ and
    $z=40\grad$.}
  \label{fig3}
 \end{center}
\end{figure}

\section{Models for FAST and feeds}

FAST is a primary-focus radio telescope. First we describe model
of the feed. For our calculation, we use the same universal
coaxial horn feed as that in \citet{dh13} with seven corrugated
walls (Fig.~\ref{fig2a}), which have good symmetry in its broad
radiation patterns (Fig.~\ref{fig2b}), very low-level sidelobes
and low-level cross polarization within the illumination angle.
We take the radiation patterns with three possible edge tapers
of $T_e$=$-$9.6~dB, $T_e$=$-$10.7~dB and $T_e$=$-$12~dB by slightly
changing the flare angle of the corrugated walls (Fig.~\ref{fig2c}).

The main reflector of FAST consists of 4400 triangular spherical
panels. The feed illuminates only some of the panels during
observations. We construct the three deformed FAST models for
observations at zenith angles of $z = 0\grad$, $z = 27\grad$ and
$z = 40\grad$ (Fig.~\ref{fig3}), exactly the same as those in \citet{dh13}.
The spherical panels in these models are assumed to be perfect
electrical conductors with no thickness. They are adjusted to
form a paraboloid with an aperture of 300~m in diameter \citep{qiu98}
during observations. In the FAST model, we include two additional
parameters as variables: the panel curvature radius $\rho_c$ and
the focal offset $\Delta f$. In the following we will calculate
the beam patterns and telescope gains for various $\rho_c$ and
$\Delta f$ to achieve the best performance.

\begin{figure}[!htb]
\begin{center}
    \subfigure[Reflected rays of the 300-m paraboloid in the FAST]{
    \includegraphics[bb=1 169 588 645,width=0.46\textwidth,height=0.35\textwidth,clip]{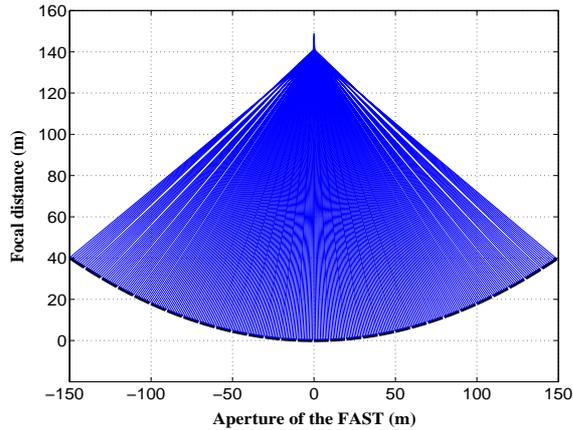}
    \label{fig4a}}
  \subfigure[zoomed view of the focal region]{
    \includegraphics[bb=0 187 568 627,width=0.46\textwidth,height=0.35\textwidth,clip]{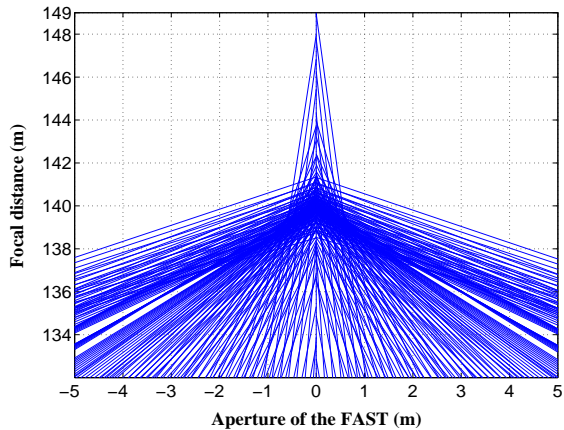}
    \label{fig4b}}
     \caption{Reflected rays of an adapted paraboloid using the
     spherical panels for the 300~m aperture of the FAST.}
  \label{fig4}
\end{center}
\end{figure}

\begin{figure}[!htb]
\begin{center}
\includegraphics[bb=0 217 584 614,width=0.46\textwidth,height=0.3\textwidth,clip]{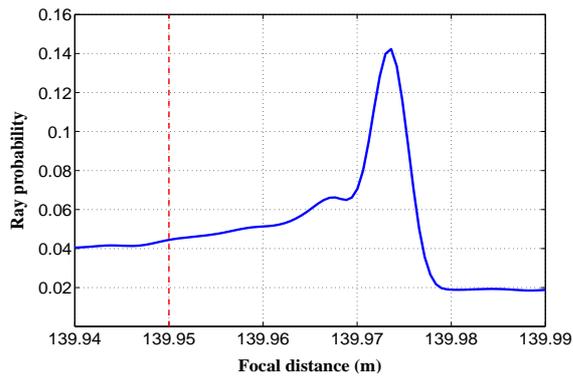}
\caption[]{The probability of reflected rays passing through the
  region near the official focal distance of 139.95~m.}
\label{fig5}
\end{center}
\end{figure}

\begin{figure}[!htb]
\begin{center}
\includegraphics[bb=14 16 750
  545,width=0.45\textwidth,height=0.32\textwidth,clip]{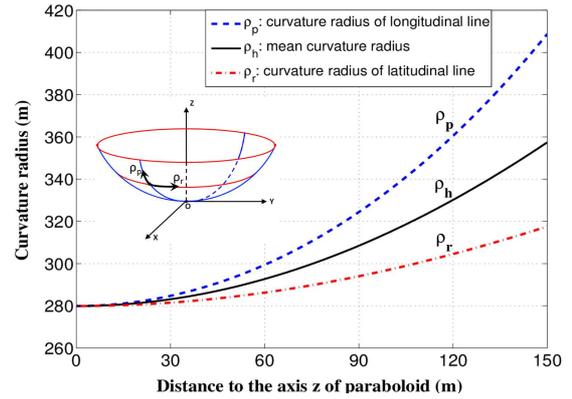}
\caption[]{Variation of curvature radius of a 300-m paraboloid with
  $f/D = 0.4665$.}
\label{fig6}
\end{center}
\end{figure}

\section{Optimisations and results}

In principle, the best focal position of FAST can be roughly
determined by geometric drawings. We demonstrate this by plotting
the reflected rays in Fig.~\ref{fig4} for a one dimensional
``parabolic line'' which has 30 arcs as cuts of the spherical
panels. The two ends of each arc are on a parabolic line. Ten
equally-spaced incident parallel rays are directed onto each
arc in Fig.~\ref{fig4a} and the zoomed view of the focal region
is shown in Fig.~\ref{fig4b}.  The best focal position should
have the highest probability for the reflected rays crossing.
The number of rays varies with respect to the focal distance,
as shown in Fig.~\ref{fig5}. With 50000 equally-spaced
incident rays directed onto each arc, the peak is slightly
offset from the exact official focal distance, which is 139.95~m
for the official focal ratio of $f/D = 0.4665$.

Now we consider the best curvature radius of the spherical panels.
For a 300-m paraboloid, the curvature of a parabolic surface varies
significantly with distance to the axis of the paraboloid (see
Fig.~\ref{fig6}). At each point the curvatures along the longitudinal
and latitudinal lines are different. The curvature radius is about
$\rho_p= \rho_r= 280$~m near the centre, while it increases to
$\rho_r= 318$~m for the latitudinal line and $\rho_p= 409$~m for the
longitudinal line at the aperture radius of 150~m. Because FAST uses
spherical panels of one given curvature radius of $\rho_c=300$~m
\citep{qiu98} or $\rho_c=318.5$~m \citep{gj10,nan11} to approximate
a paraboloid, there must be very different deviations in the central
and outer part of the mimic paraboloid due to the inconsistent
curvature radii on an ideal paraboloid. Without consideration of the
illumination function of a practical feed, the minimum deviation for
FAST's 300~m aperture was found for the spherical panels having a
curvature radius of 318~m \citep{gj10}.

\subsection{Beam optimisation via panel curvature and focal offset}

We now calculate the beam patterns and the gains by using the three
feed patterns and the FAST models shown in Sect.~2 with the Shooting
and Bouncing Ray method for electromagnetic computations. We try to
get the best calculation results for the parameter space of $\rho_c$
and $\Delta f$ in $280<\rho_c<350$~m and $-1< \Delta f <9$~cm. The
steps for the two parameters are 5~m for $\rho_c$ and 1~cm for $\Delta f$,
with interpolations for some specific known values if necessary. In our
calculation, each spherical panel is meshed by at least 1500 curved
triangular elements, and the smallest element in critical areas has
an edge length of 0.4~mm. A total number of $10^8$ rays are directed
from the feed with a ray distance of 0.1$\sim$50~mm on the spherical
panel after ``adaptive ray sampling''. Experiments show that the
accuracy of these settings is high enough to detect small deviations
from the spherical panels and the focal shift in FAST. Moreover,
reflections from the feed cabin are ignored in our calculation since
it might affect the optimal focal distance by reflecting rays into
the backlobe of the feed.

\begin{figure*}[!hbt]
\begin{center}
\includegraphics[bb=0 190 561 637,width=0.325\textwidth,height=0.25\textwidth,clip]{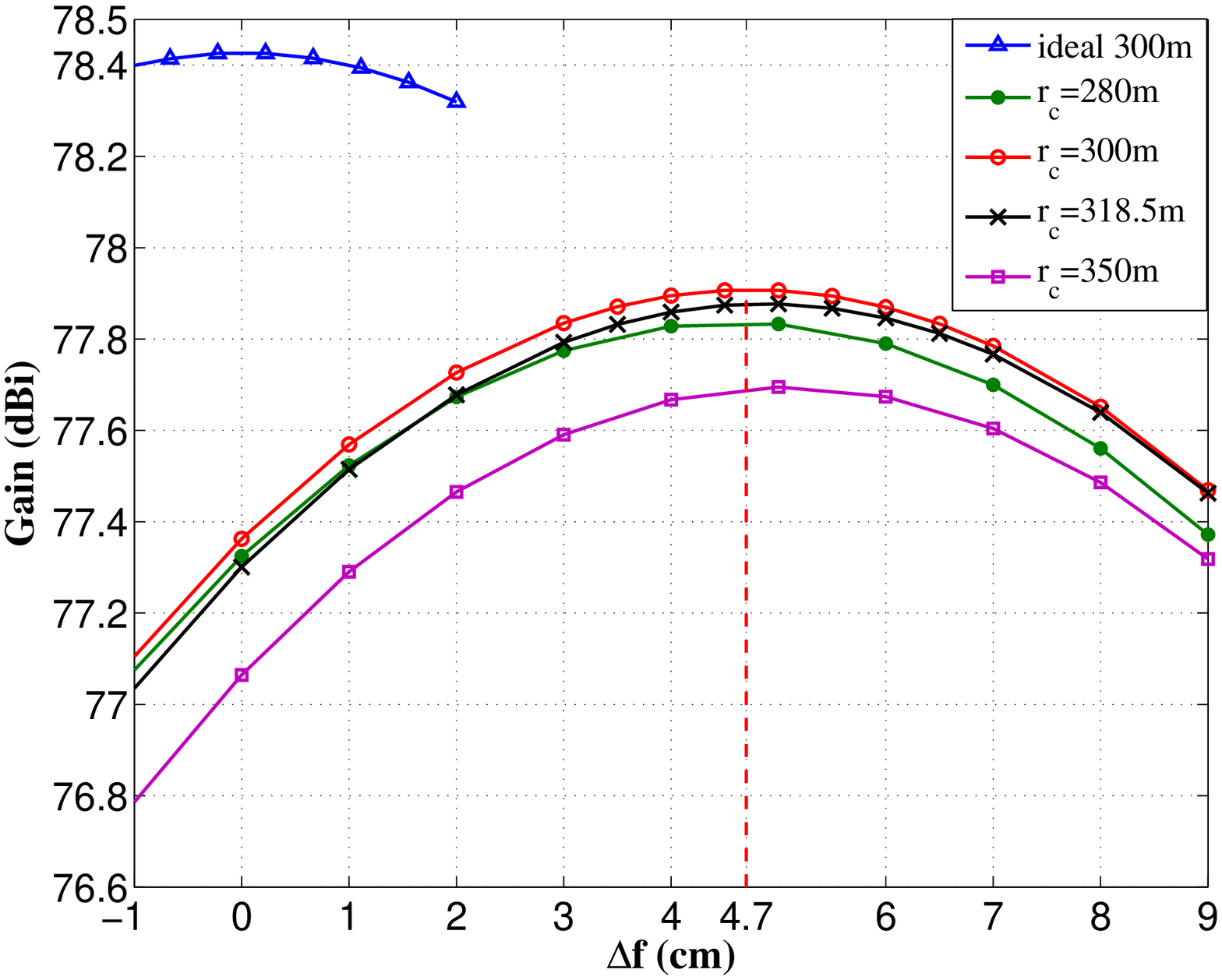}
\hspace{-0.25mm}   \includegraphics[bb=22 188 560 636,width=0.315\textwidth,height=0.25\textwidth,clip]{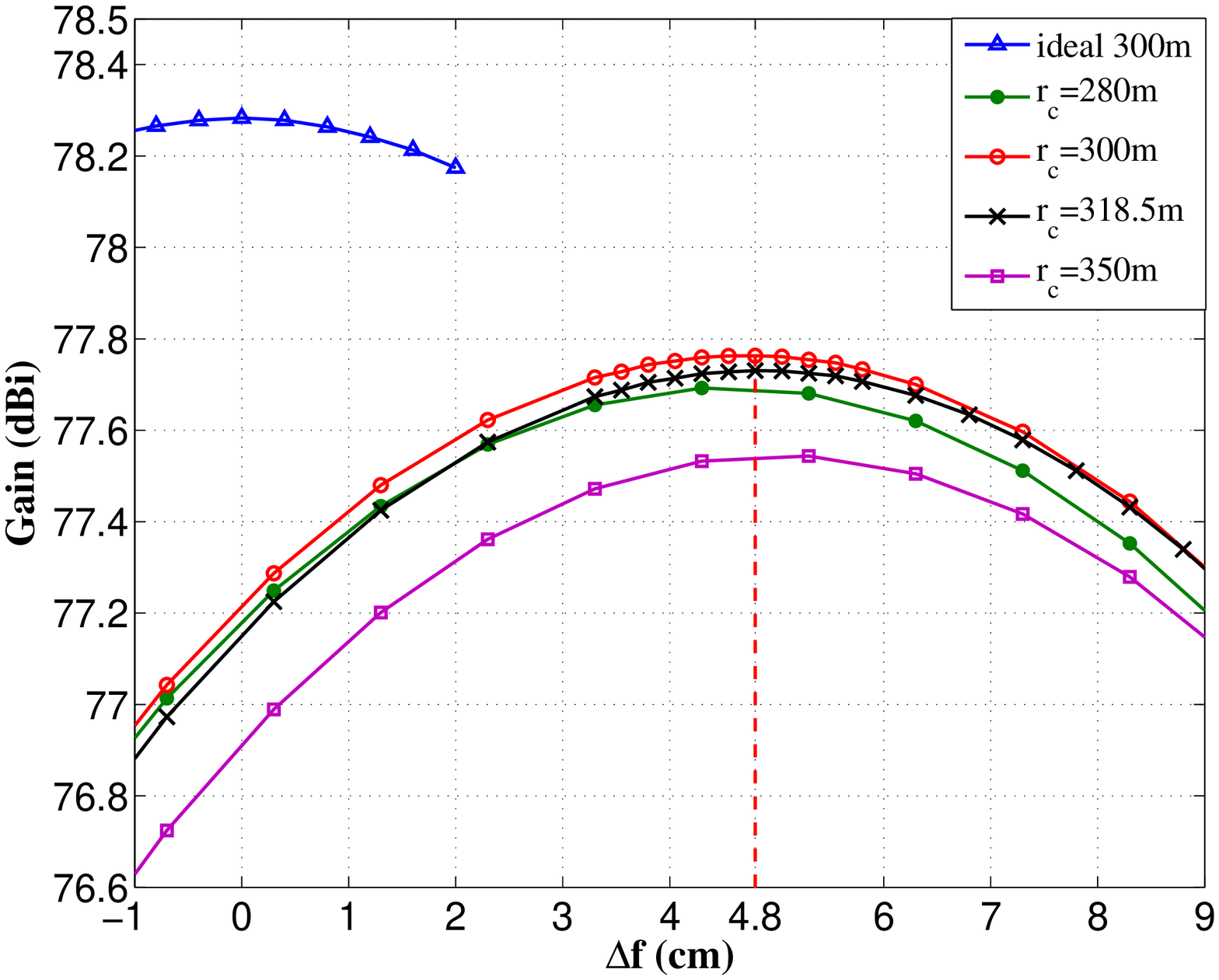}
\includegraphics[bb=19 188 560 636,width=0.315\textwidth,height=0.25\textwidth,clip]{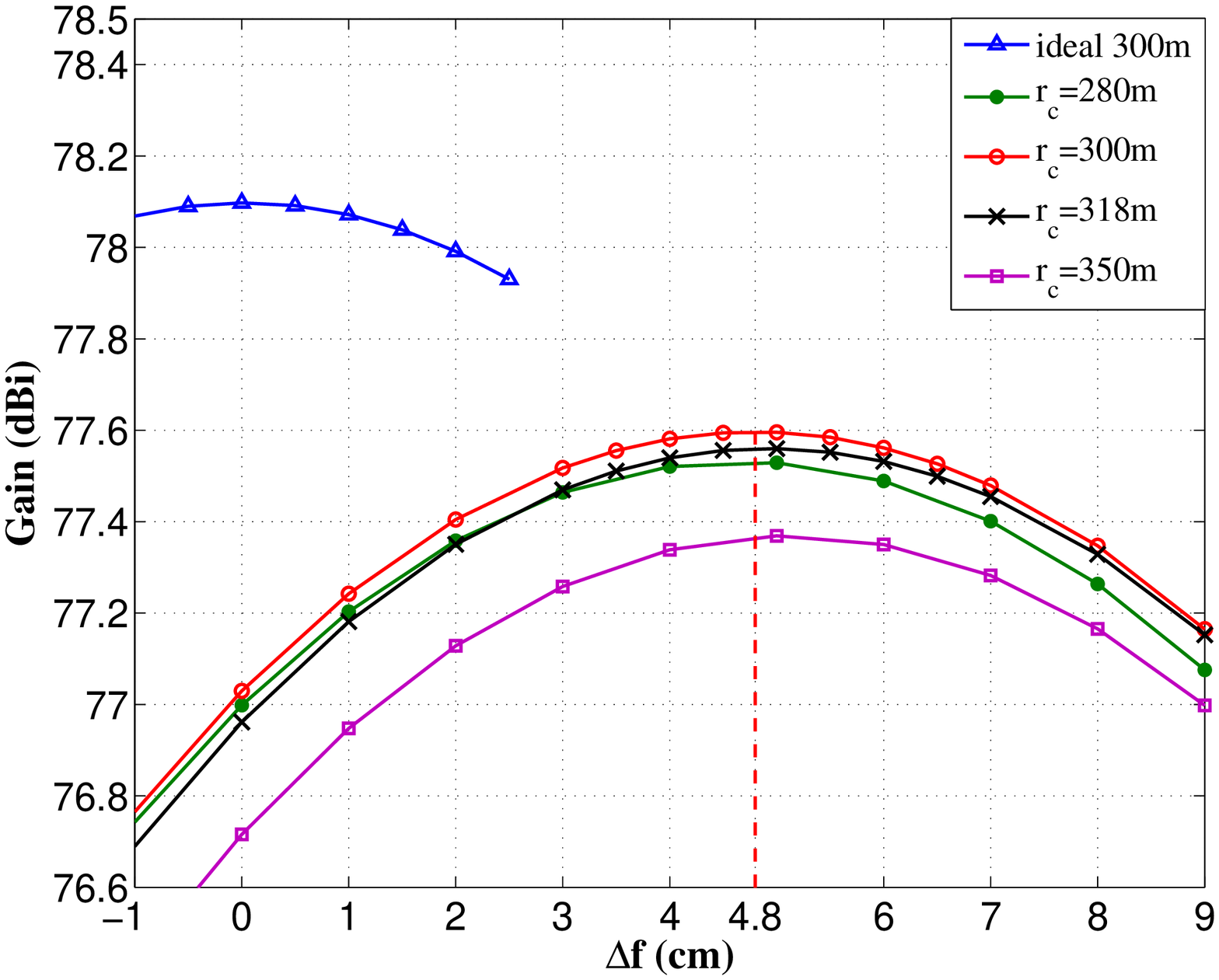}\\[2mm]

\subfigure[$T_{e}$=$-$9.6 dB]{
\hspace{-2mm}\includegraphics[bb=2 185 506 639,width=0.3275\textwidth,height=0.275\textwidth,clip]{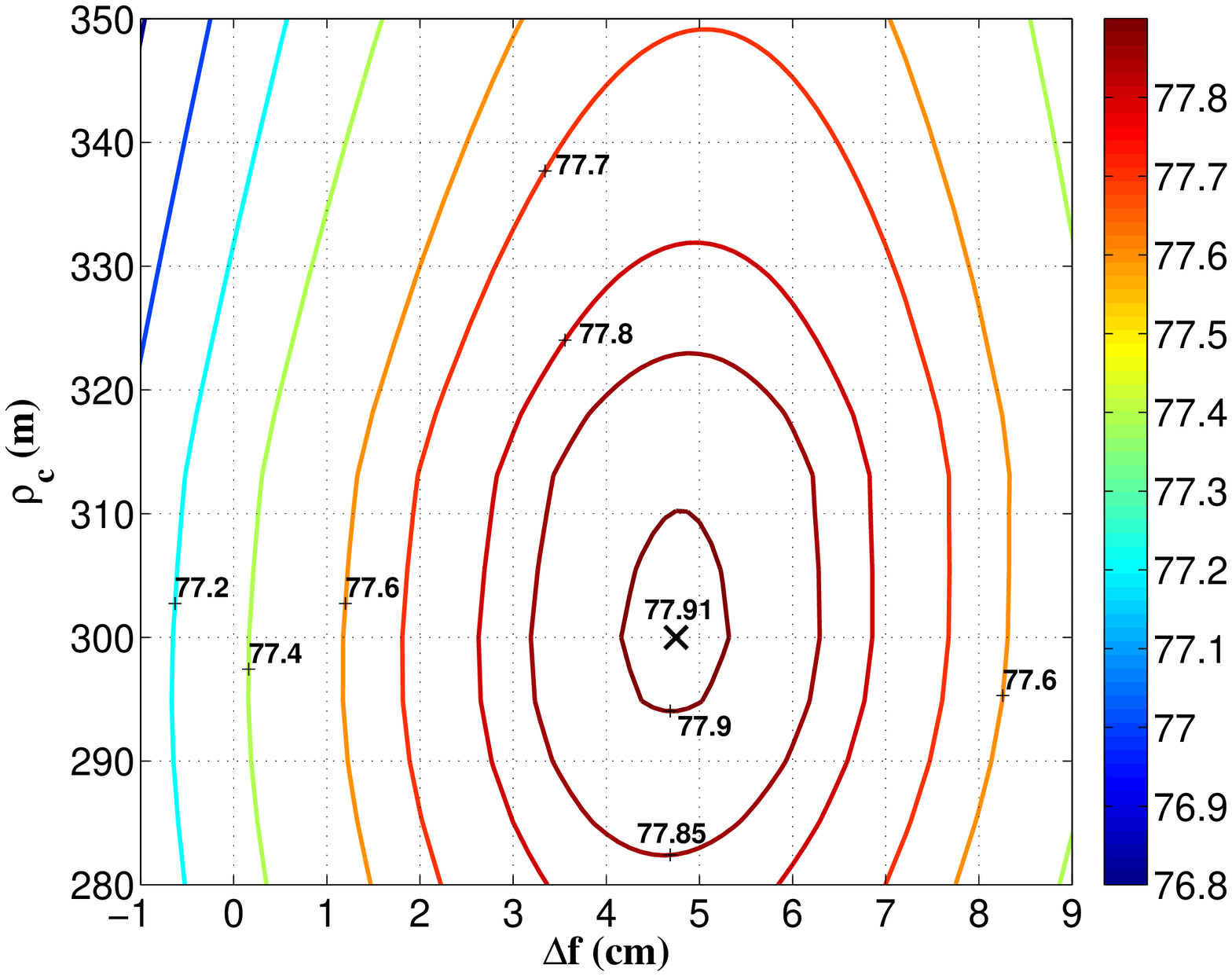}}
\subfigure[$T_{e}$=$-$10.7 dB]{
   \includegraphics[bb=34 186 505 639,width=0.3105\textwidth,height=0.275\textwidth,clip]{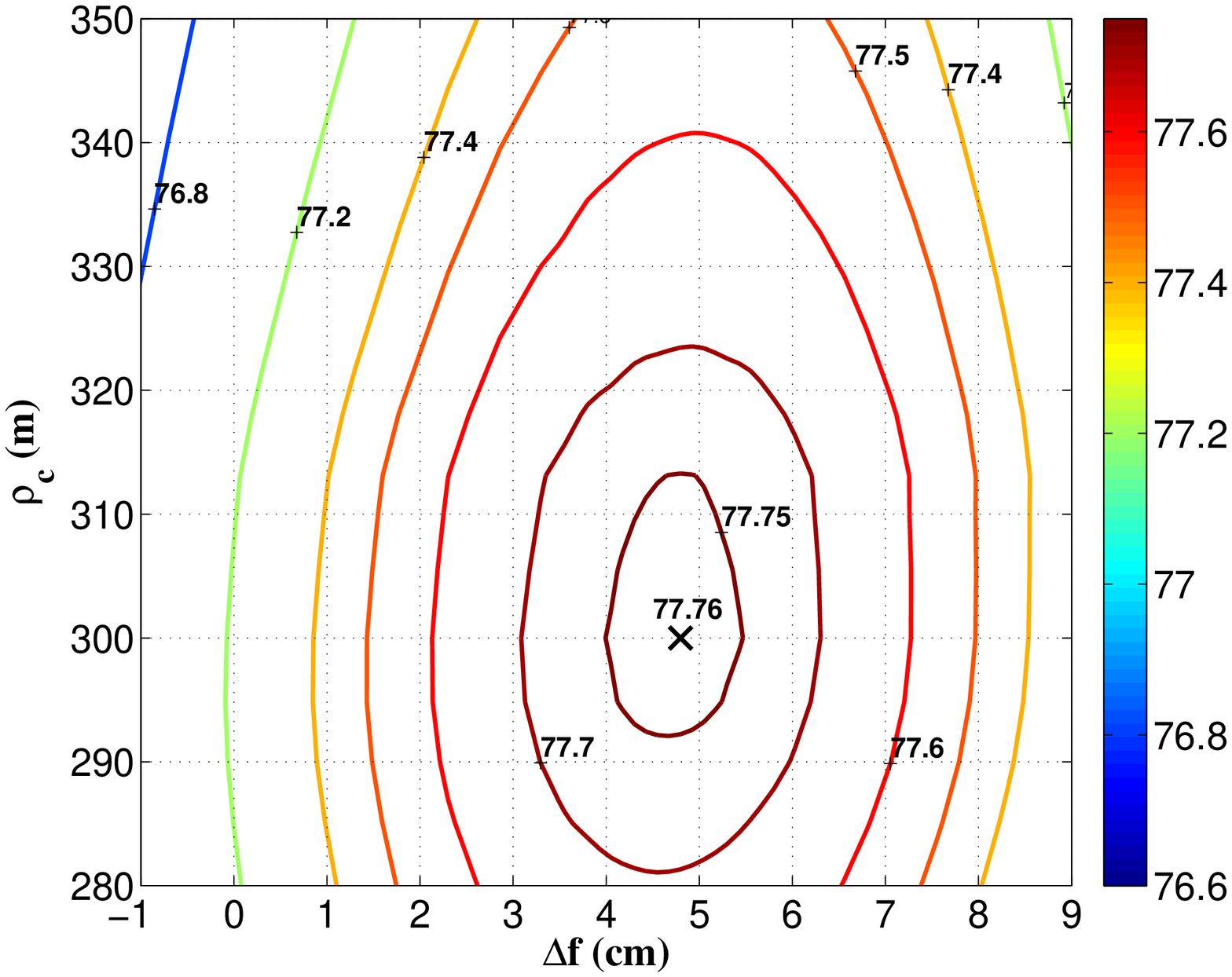}}
\subfigure[$T_{e}$=$-$12 dB]{
   \includegraphics[bb=33 186 507 640,width=0.315\textwidth,height=0.275\textwidth,clip]{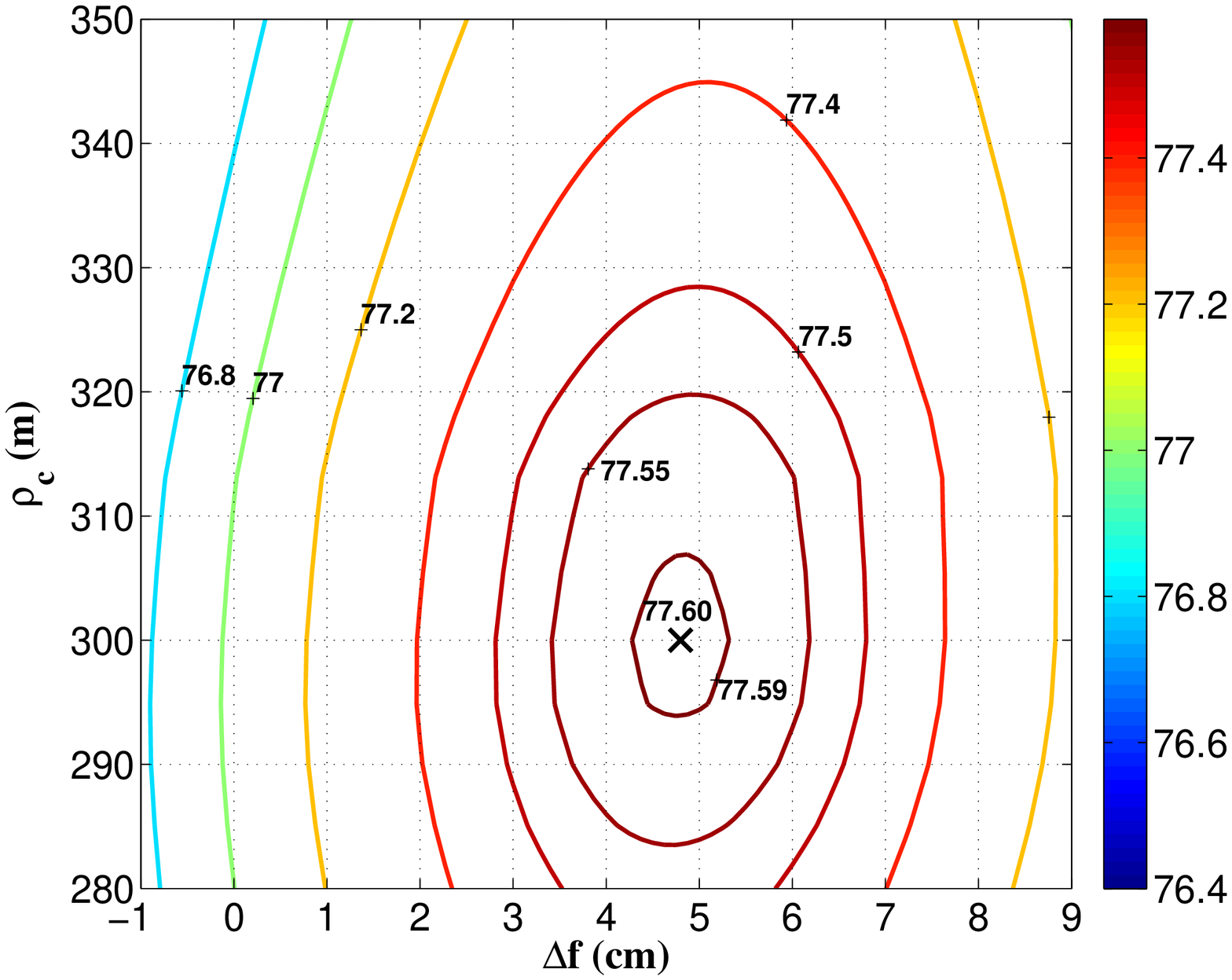}}

  \caption{Curves for the gain as a function of $\Delta f$ for
    $\rho_c$=280~m, 300~m, 318.5~m and 350~m in {\it the upper panels},
    and gain contours for $\Delta f$ and $\rho_c$ in {\it the lower
      panels} for three different edge tapers of the feed:
    $T_e=-9.6$~dB, $T_e=-10.7$~dB and $T_e=-12.0$~dB. For comparison,
    the gain for the ideal 300-m paraboloid as a function of the focal
    shift $\Delta f$ is also calculated and plotted as blue lines in
    {\it the upper panels}.}
  \label{fig7}
\end{center}
\end{figure*}
\begin{figure*}[!hbt]
\begin{center}
{ Before Optimisation}\\
\includegraphics[bb=12 186 540 632,width=0.33\textwidth,height=0.285\textwidth,clip]{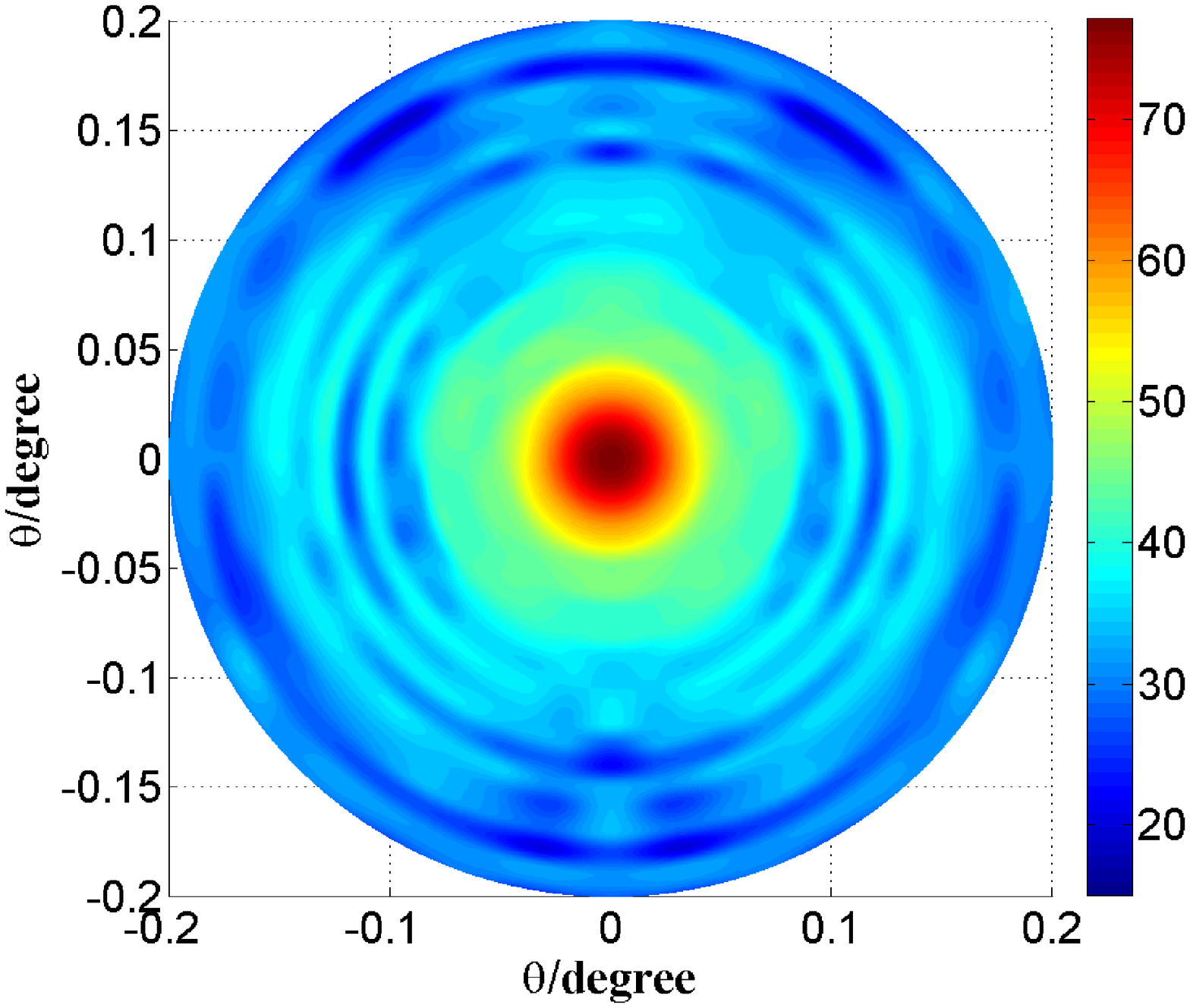}
\includegraphics[bb=38 183 540 632,width=0.316\textwidth,height=0.285\textwidth,clip]{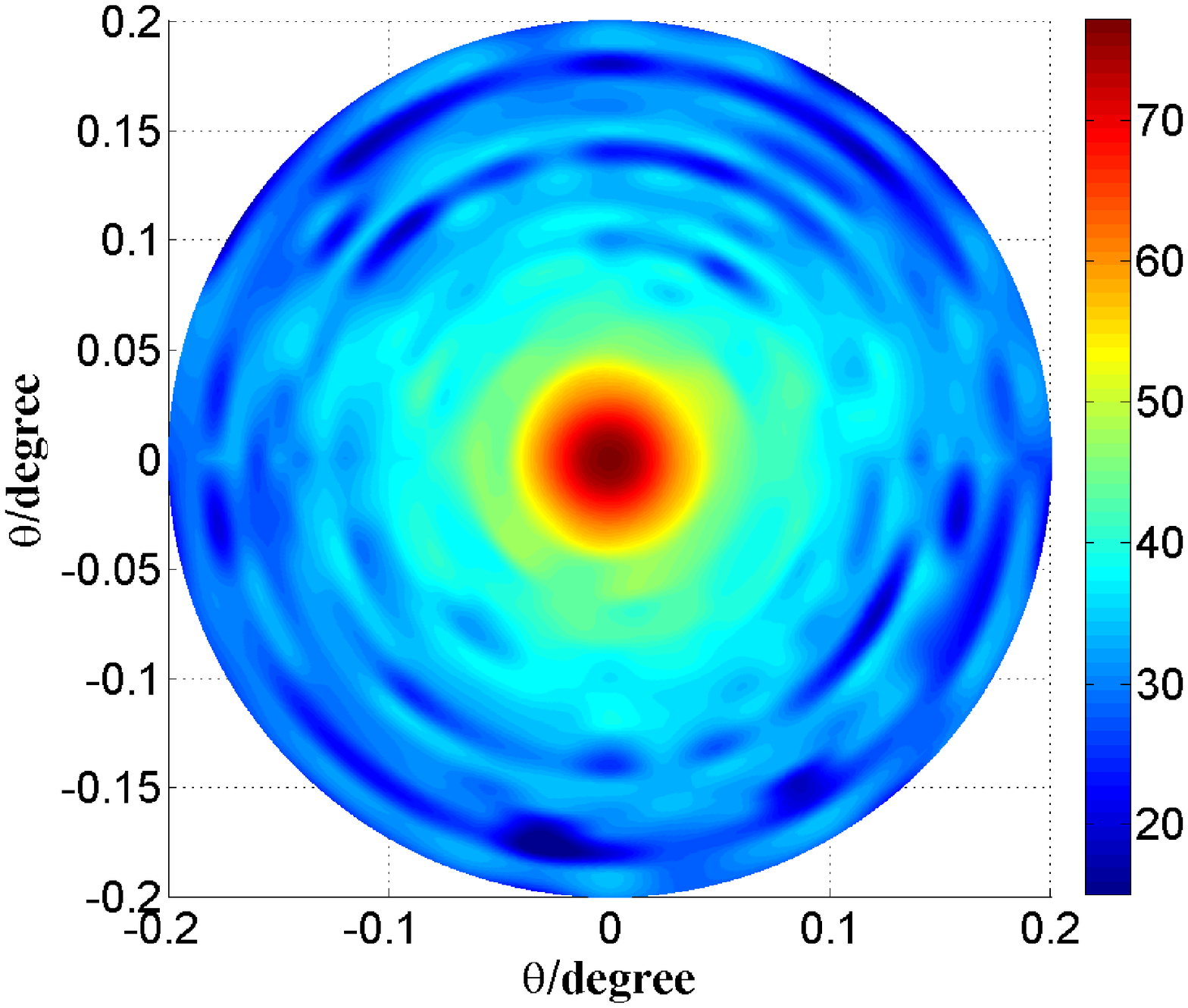}
\includegraphics[bb=38 183 540 632,width=0.316\textwidth,height=0.285\textwidth,clip]{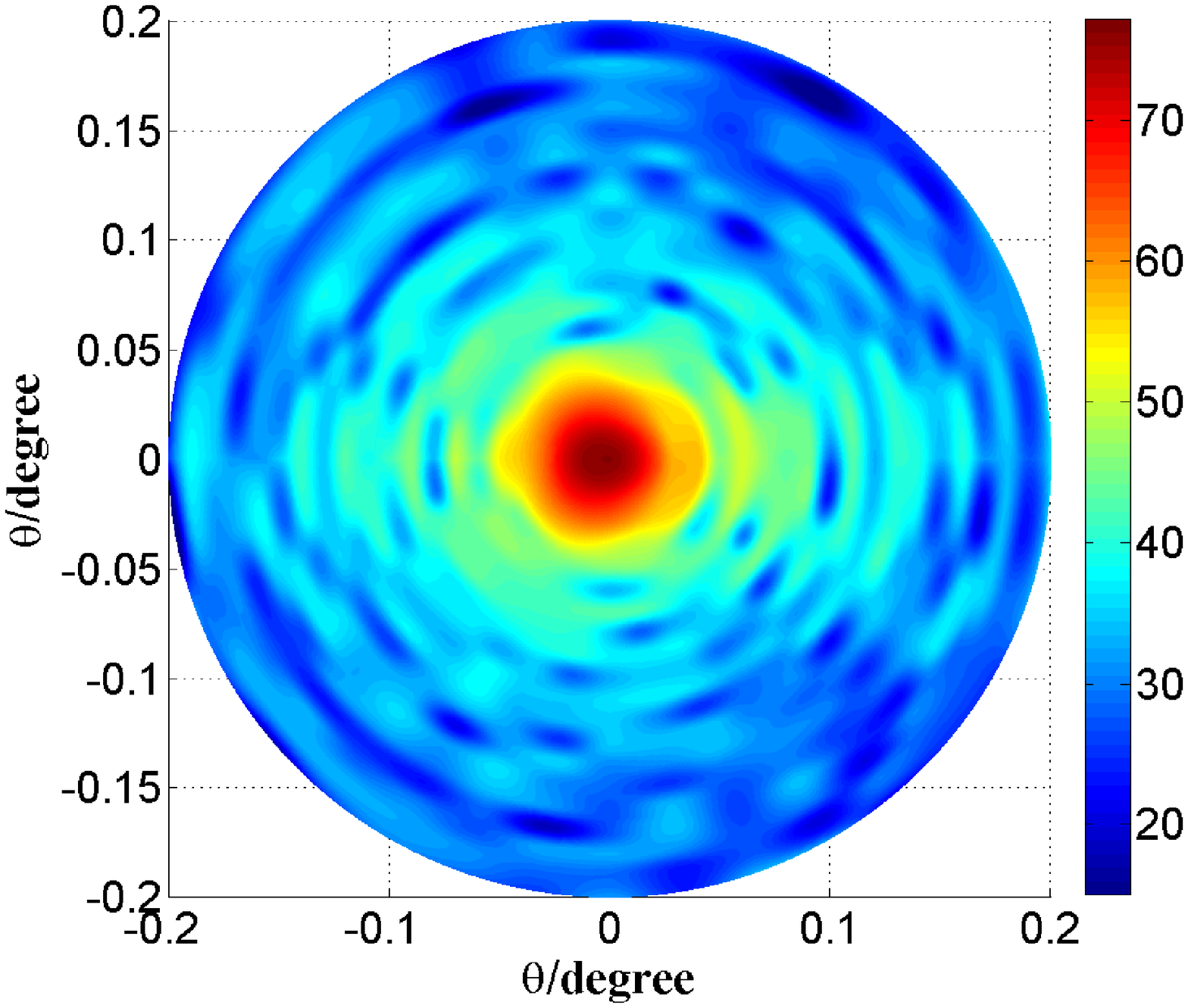}\\
{ After Optimisation via panel curvature and focal offset}\\
\includegraphics[bb=13 178 539 633,width=0.33\textwidth,height=0.2825\textwidth,clip]{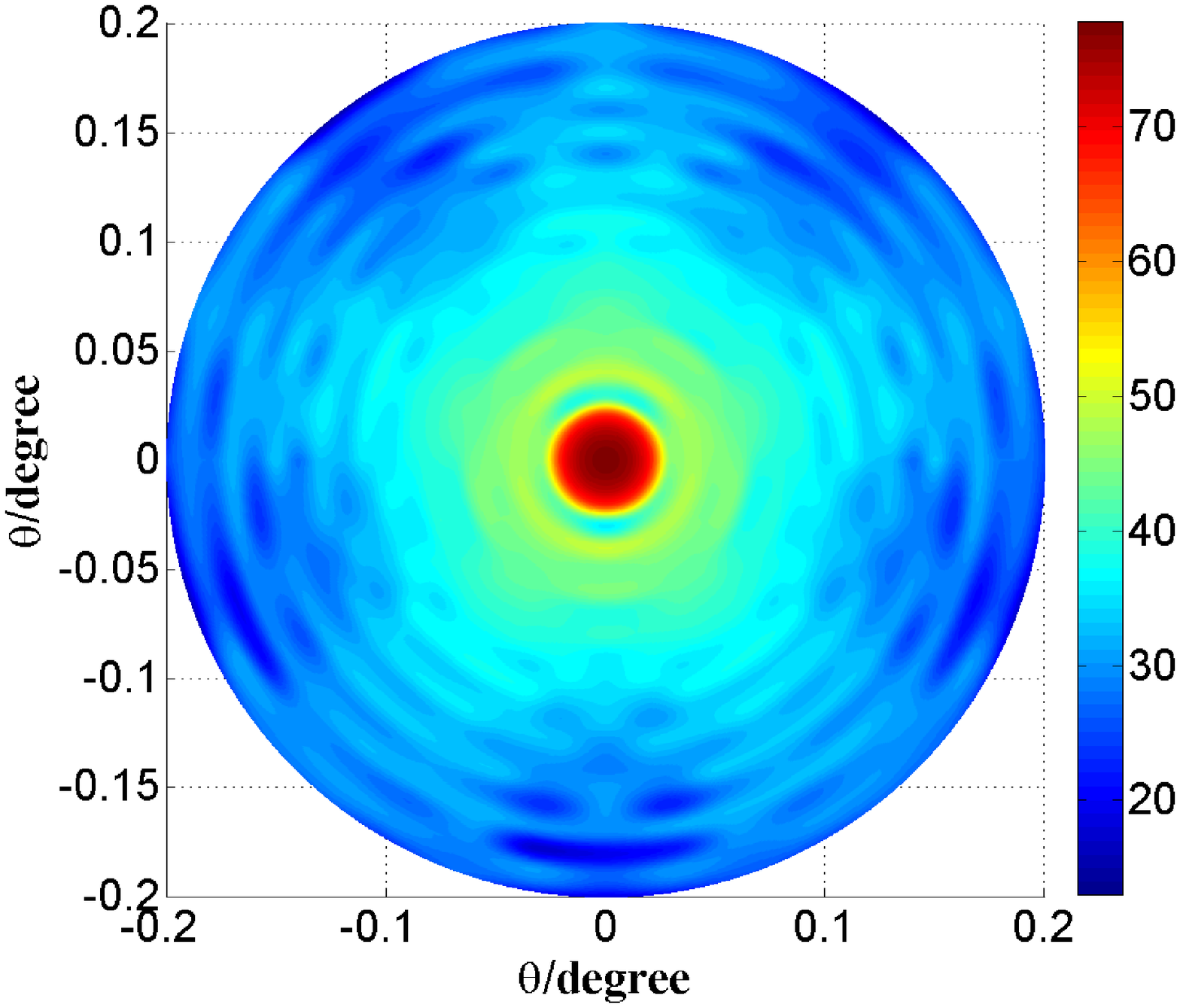}
\includegraphics[bb=36 180 539 634,width=0.316\textwidth,height=0.2825\textwidth,clip]{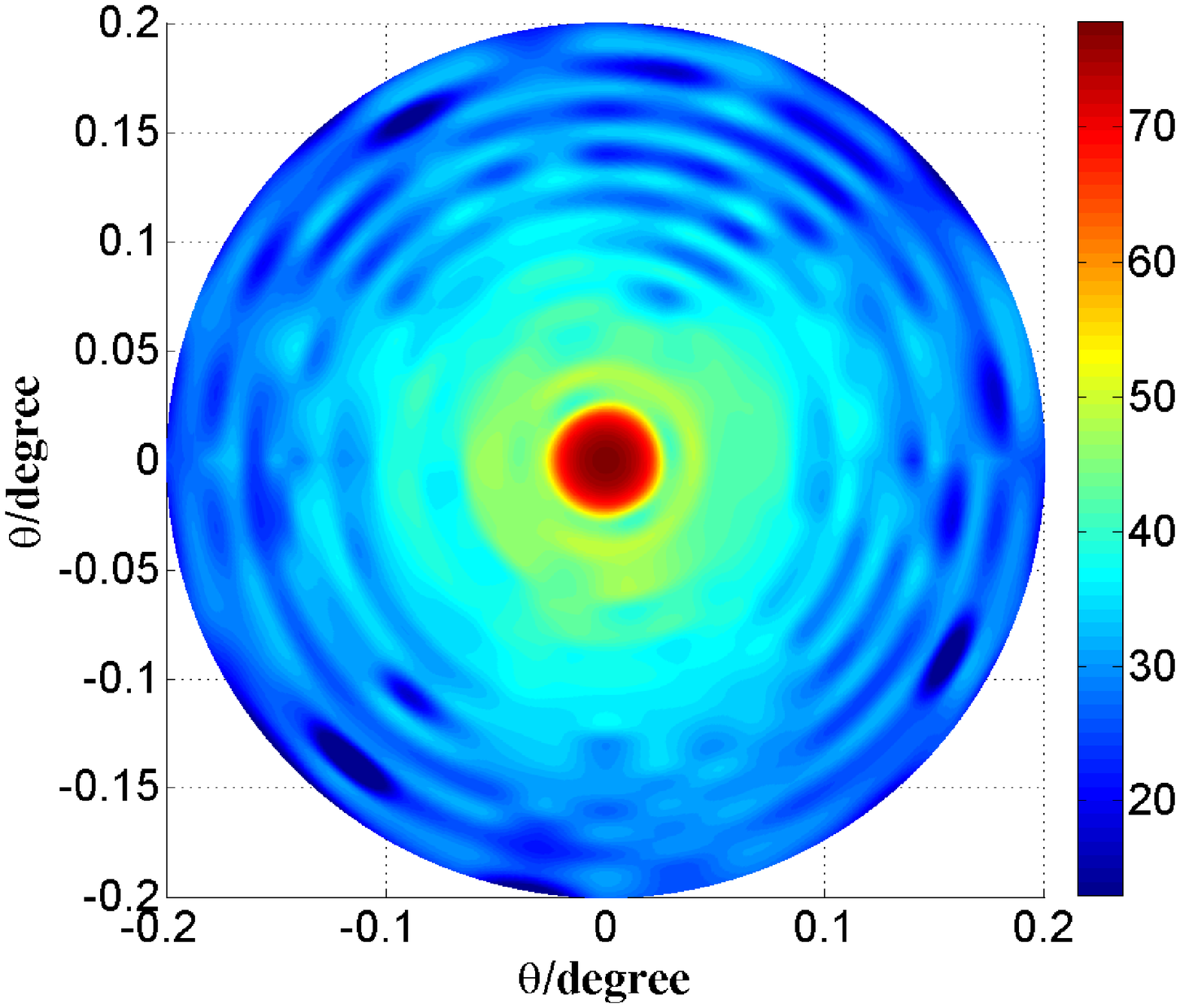}
\includegraphics[bb=43 186 540 625,width=0.316\textwidth,height=0.2825\textwidth,clip]{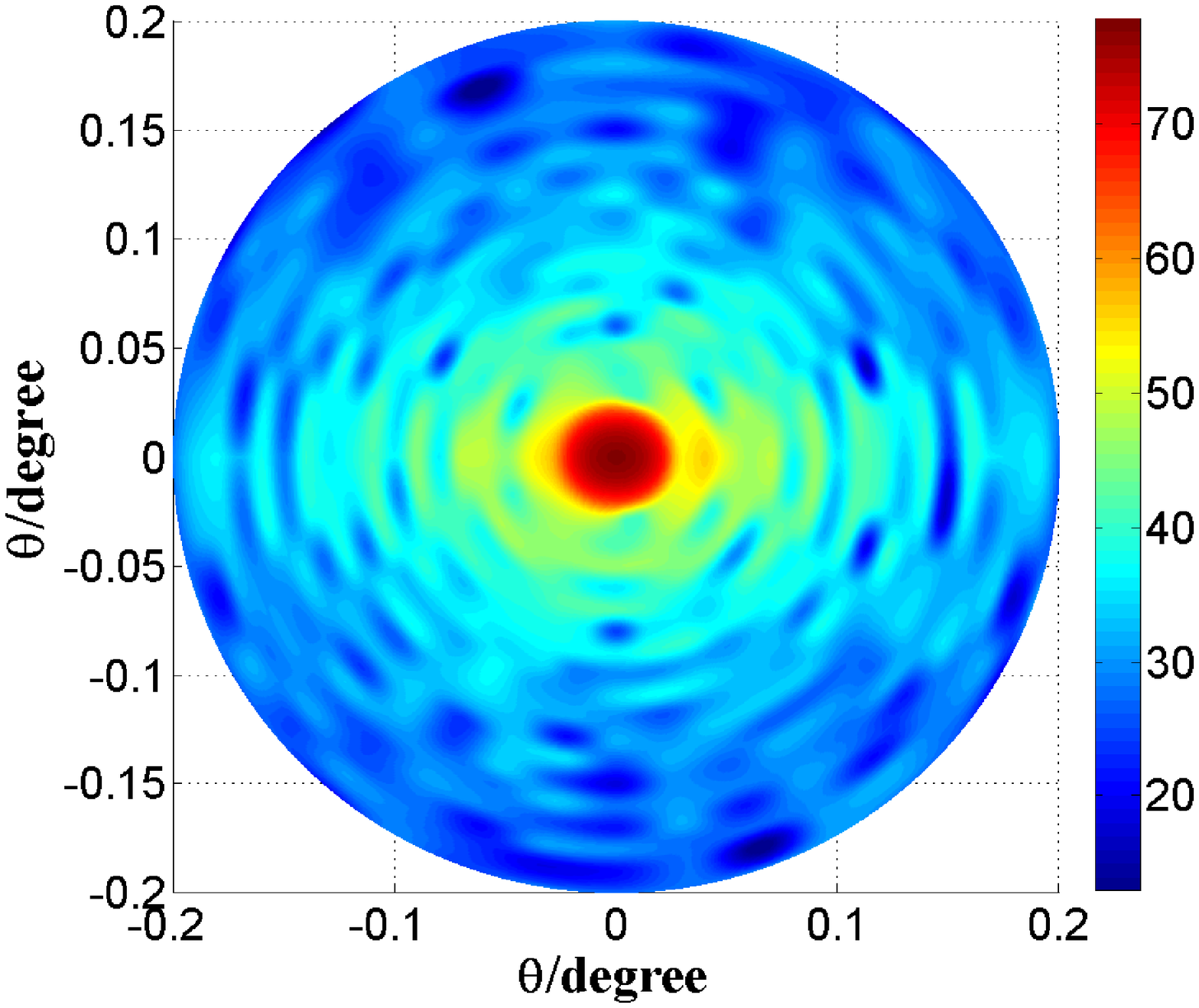}\\
{ After Optimisation via panel positioning}\\
\includegraphics[bb=42 39 517 432,width=0.33\textwidth,height=0.2825\textwidth,clip]{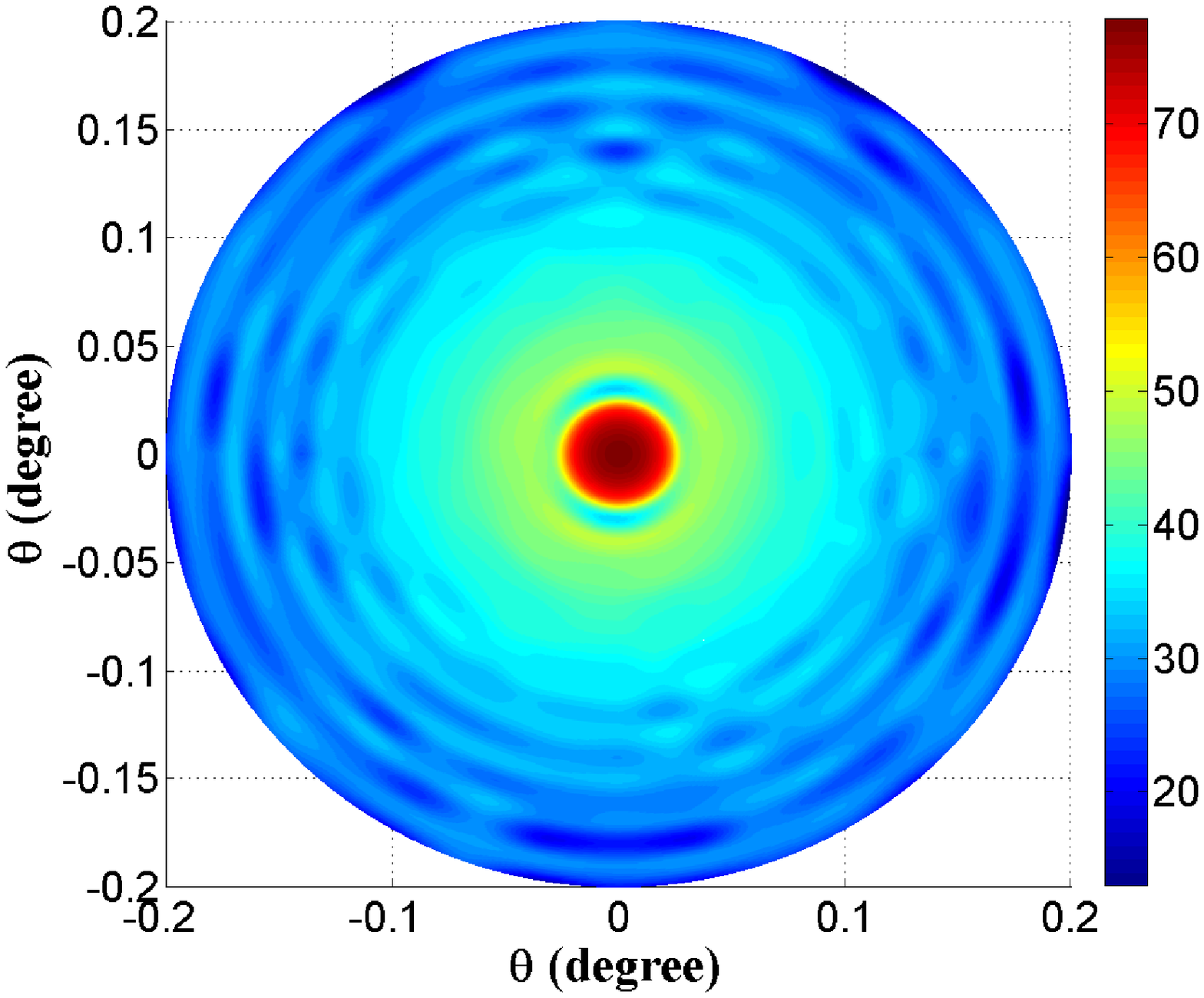}
\includegraphics[bb=37 186 543 627,width=0.316\textwidth,height=0.2825\textwidth,clip]{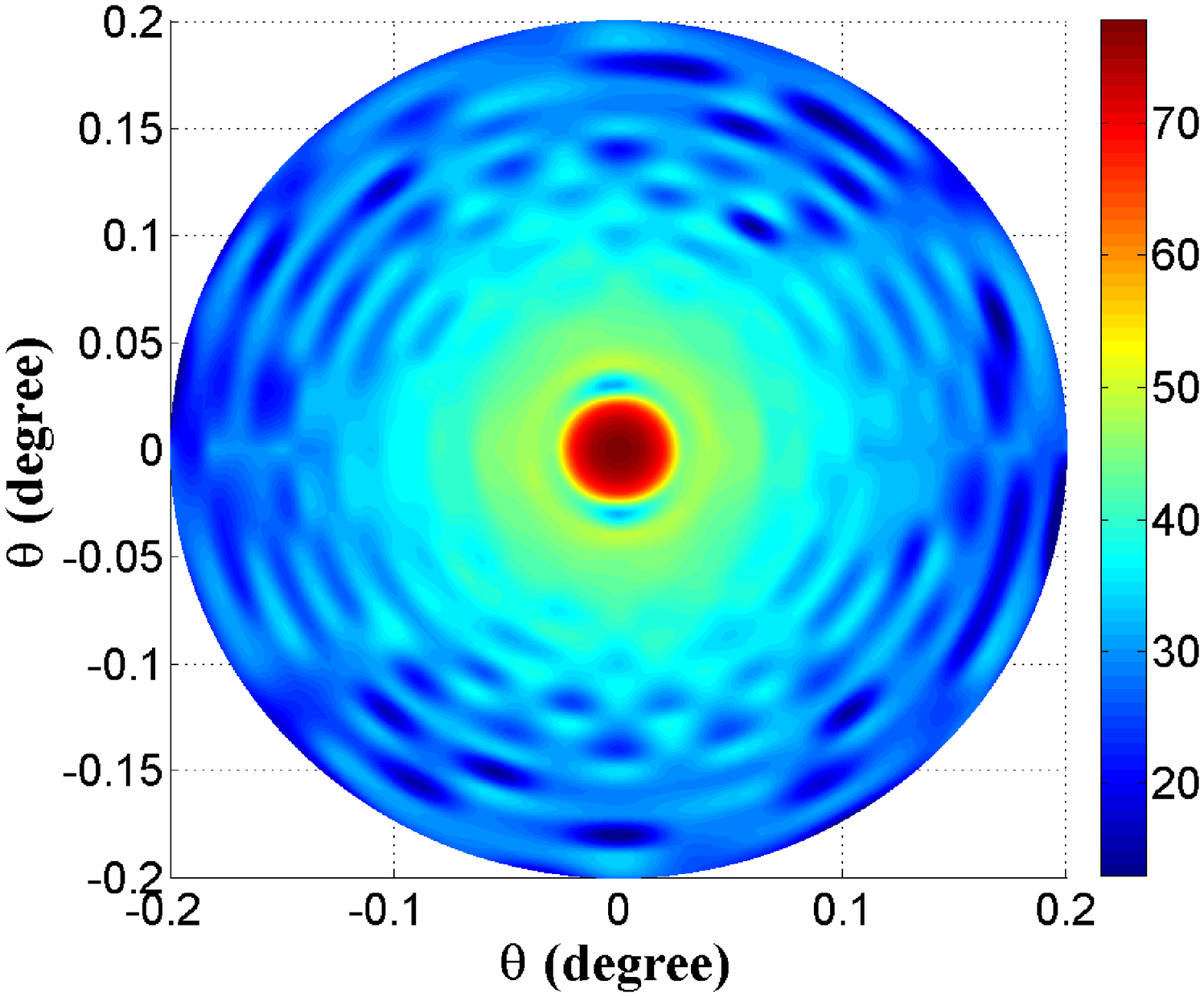}
\includegraphics[bb=48 195 531 622,width=0.316\textwidth,height=0.2825\textwidth,clip]{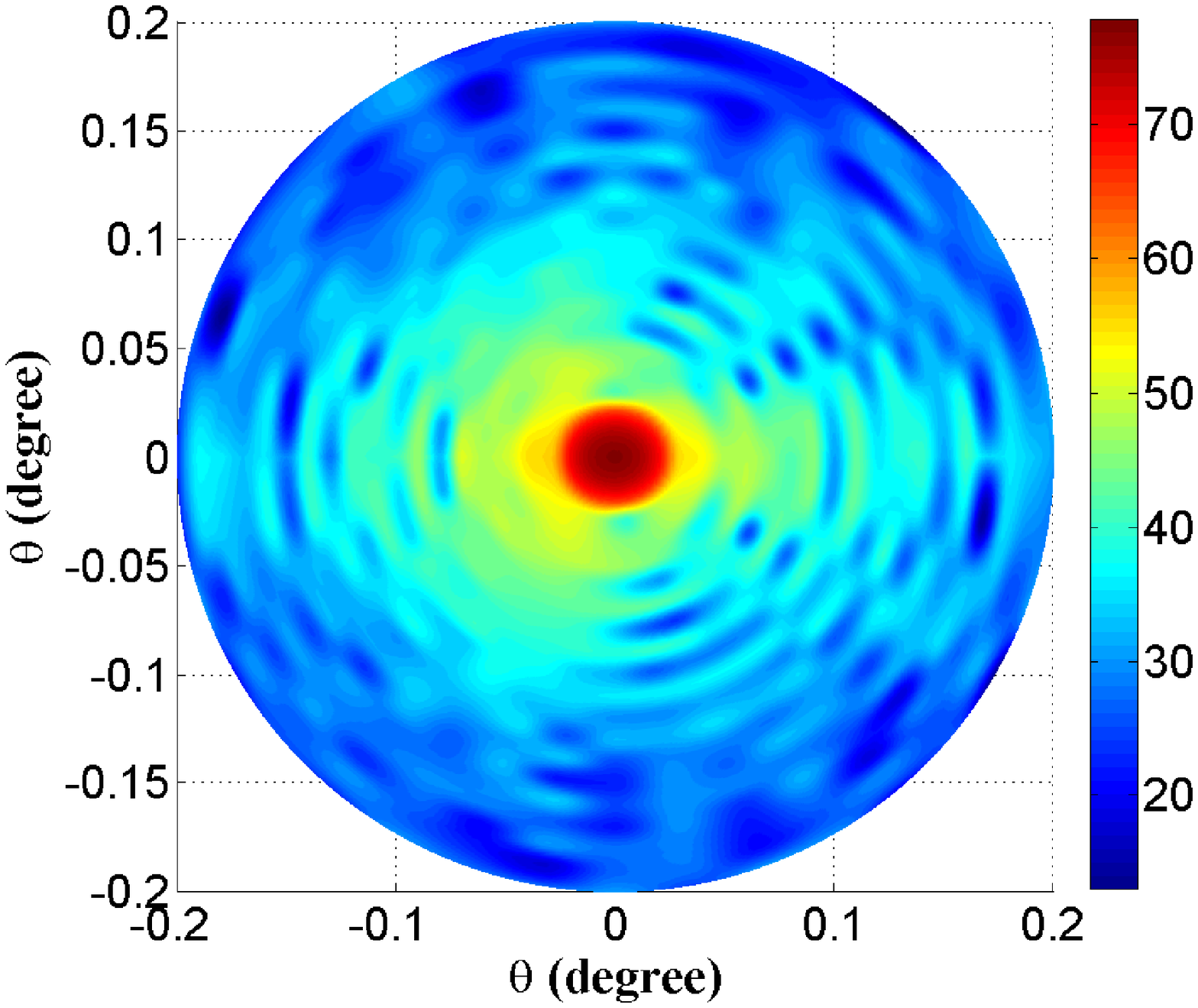}\\
  \subfigure[$z=0\grad$]{
\includegraphics[bb=0 194 588 621,width=0.33\textwidth,height=0.225\textwidth,clip]{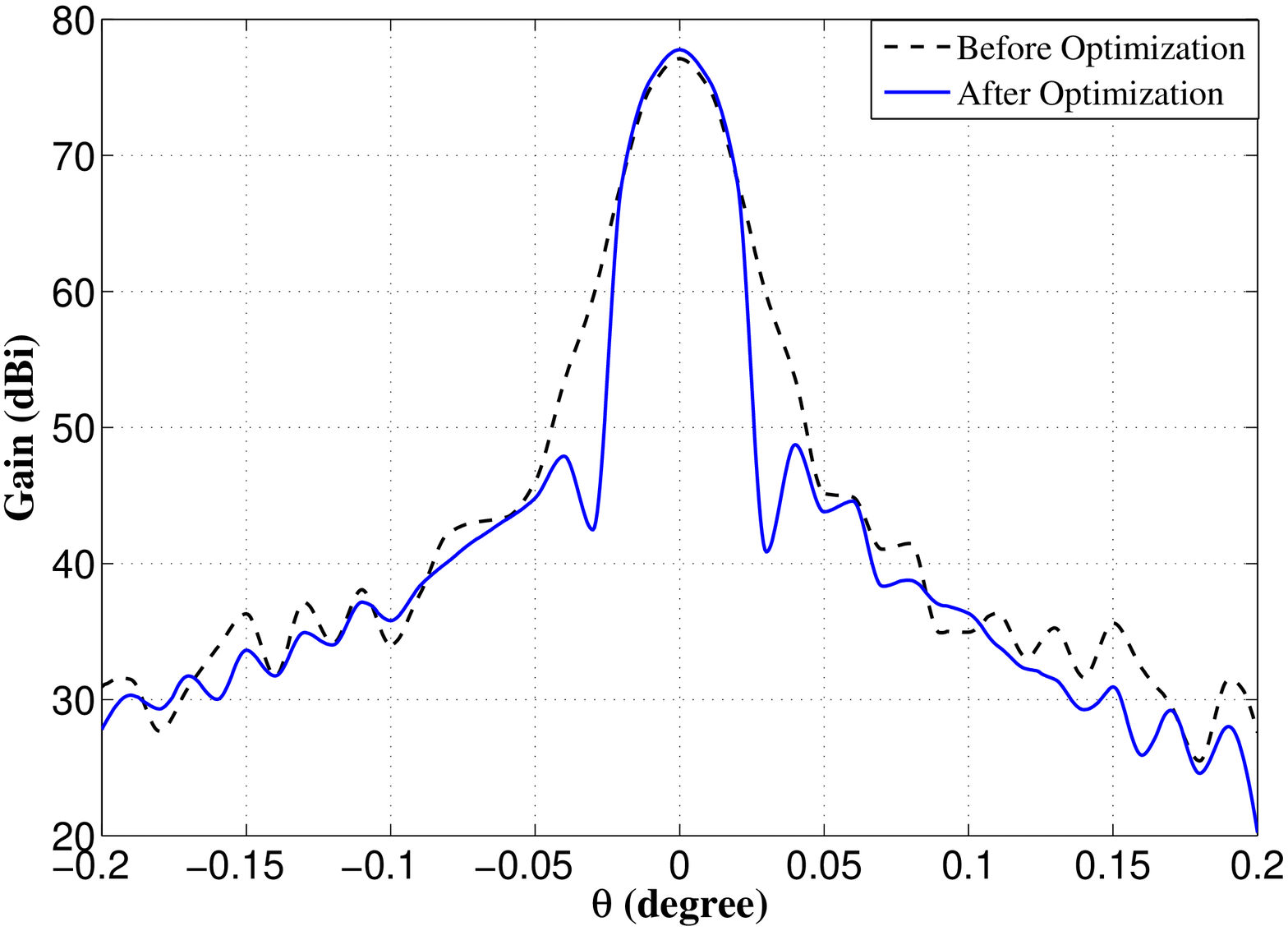}}
  \subfigure[$z=27\grad$]{
\includegraphics[bb=23 194 588 621,width=0.316\textwidth,height=0.225\textwidth,clip]{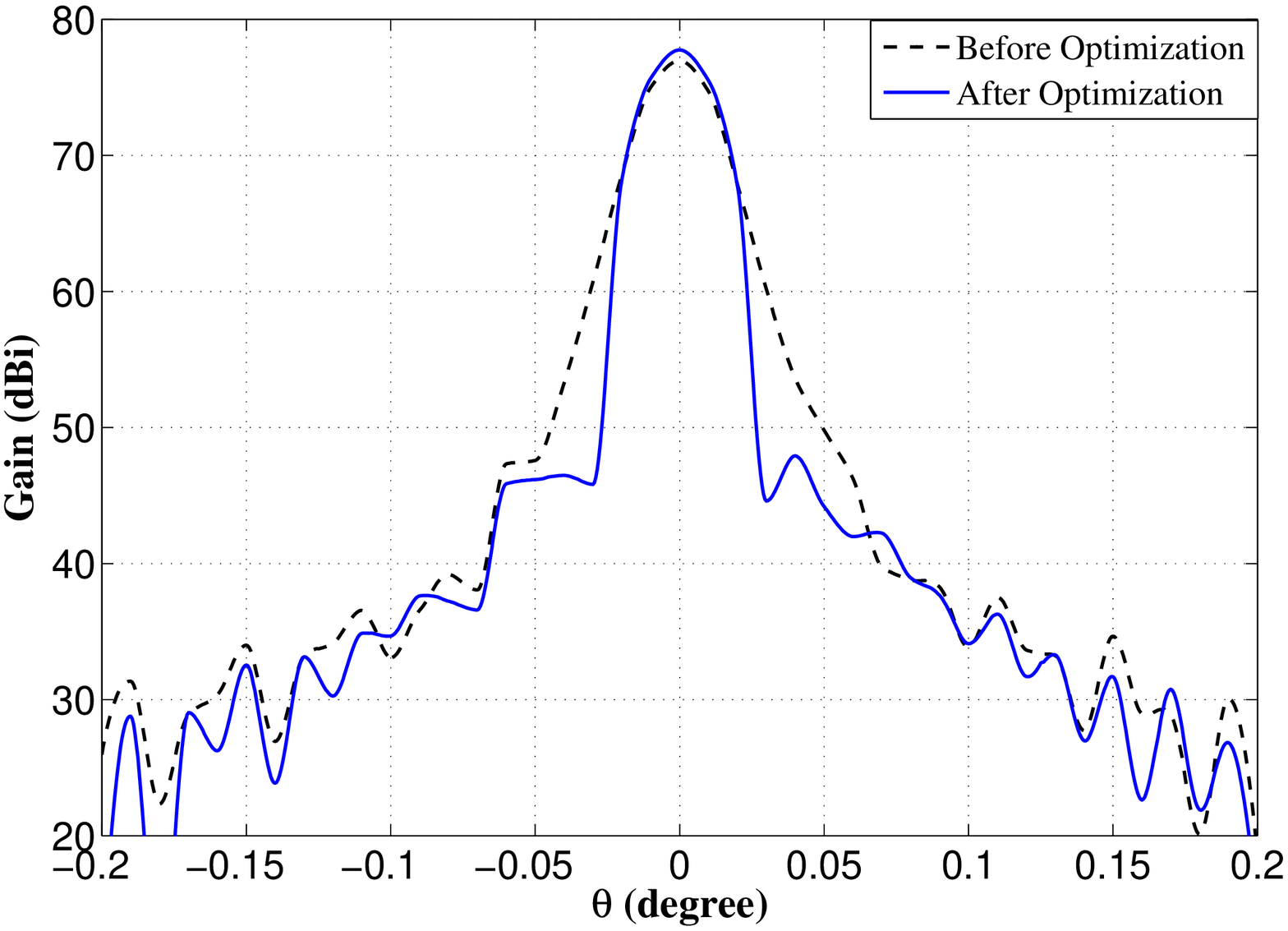}}
  \subfigure[$z=40\grad$]{
\includegraphics[bb=60 236 580 605,width=0.316\textwidth,height=0.225\textwidth,clip]{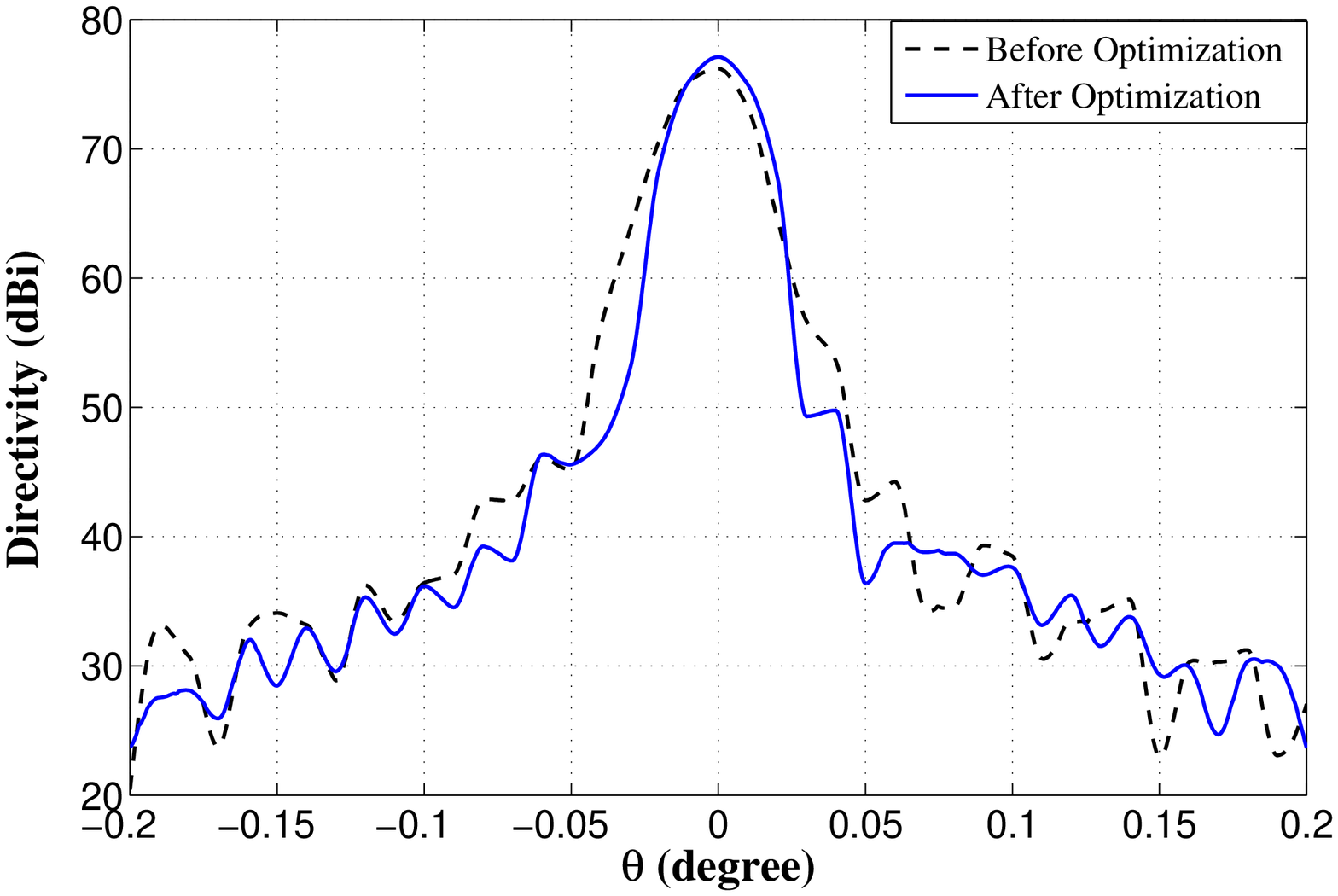}}\\
\caption{The beam patterns of the FAST at 3 GHz for observations at
  $z=0\grad$, $27\grad$ and $40\grad$ with a feed of $T_{e}$=$-$10.7
  dB. Cuts in the $\phi=45\grad$ plane are shown in the last row for
  beams before and after the optimisation.}
\label{fig8}
\end{center}
\end{figure*}
\begin{table*}[!hbt]
\caption{Beam performance of the FAST at 3~GHz before and after optimisation.}
\begin{center}
\begin{tabular} {lrrrr}
\hline
\hline
Beam performance                       &300~m paraboloid           &  $z$=0\grad       &  $z$=27\grad     &  $z$=40\grad \\
\hline
\multicolumn{5}{c}{\textit{Before Optimisation}}        \\
Gain ($G$)                                         &  78.28 dBi         & 77.10 dBi         &  76.96 dBi       & 76.05 dBi \\
Cross-polarisation($\phi$=45\grad)                   & $-$49.62 dB        & $-$49.25 dB       & $-$49.74 dB      & $-$29.80 dB\\
First sidelobe                                     & $-$29.58 dB        & $-$34.61 dB       & $-$30.16 dB      & $-$24.76 dB \\
HPBW                                               & $1.41^{'}$         & $1.48^{'}$$\sim$$1.5^{'}$  & $1.41^{'}$$\sim$$1.44^{'}$  &  $1.52^{'}$$\sim$$1.62^{'}$  \\
Aperture efficiency ($\epsilon_{\mathrm ap}$)      & 75.65$\%$          & 57.55$\%$         & 55.82$\%$        & 59.29$\%$   \\
Effective diameter ($d_{eff}$)                     & 260.92 m           & 227.59 m          & 223.95 m         & 202.07 m   \\
\hline
\multicolumn{5}{c}{\textit{After Optimisation via panel curvature and focal offset}}        \\
Gain($G$)                                          &  78.28 dBi         & 77.76 dBi         &  77.75 dBi       & 77.11 dBi \\
Cross-polarisation($\phi$=45\grad)                   & $-$49.62 dB        & $-$51.61 dB       & $-$51.16 dB      & $-$30.55 dB\\
First sidelobe                                     & $-$29.58 dB        & $-$28.23 dB       & $-$28.58 dB      & $-$21.36 dB \\
HPBW                                               & $1.41^{'}$         & $1.44^{'}$        & $1.43^{'}$       & $1.35^{'}$$\sim$$1.62^{'}$  \\
Aperture efficiency($\epsilon_{\mathrm ap}$)       & 75.65$\%$     & 67.11$\%$         & 66.91$\%$        & 67.49$\%$   \\
Effective diameter($d_{eff}$)                      & 260.92 m           & 245.77 m          & 245.40 m         & 228.22 m   \\
\hline
\multicolumn{5}{c}{\textit{After Optimisation via panel positioning}}        \\
Gain($G$)                                          &  78.28 dBi         & 77.78 dBi         &  77.75 dBi       & 77.05 dBi \\
Cross-polarisation($\phi$=45\grad)                   & $-$49.62 dB        & $-$51.35 dB       & $-$51.03 dB      & $-$30.25 dB\\
First sidelobe                                     & $-$29.58 dB        & $-$28.68 dB       & $-$28.65 dB      & $-$22.25 dB \\
HPBW                                               & $1.41^{'}$         & $1.44^{'}$        & $1.44^{'}$       & $1.37^{'}$$\sim$$1.64^{'}$  \\
Aperture efficiency($\epsilon_{\mathrm ap}$)       & 75.65$\%$          & 67.36$\%$         & 66.96$\%$        & 68.37$\%$   \\
Effective diameter($d_{eff}$)                      & 260.92 m           & 246.22 m          & 245.49 m         & 226.65 m   \\

\hline
\hline
\end{tabular}
\label{tab1}
\end{center}
\end{table*}

The beam patterns and telescope gains at 3~GHz have been found to be a
function of the focal shift $\Delta f$ and curvature radius $\rho_c$
(Fig.~\ref{fig7}). The optimal values of these two parameters are
$\rho_c=300$~m and $\Delta f=4.8$~cm, which are almost the same for
different levels of the feed edge taper $T_e$. The best curvature
radius, $\rho_c = 300$~m, is smaller than 318.5~m obtained by
\citet{gj10}, because the feed illumination is considered in this
paper. A lower level of feed edge taper results in smaller best
gains. For the ideal 300~m paraboloid, phase centers of the three feed
patterns are carefully calculated and placed at the focal point
exactly, so there are almost no focal shifts and the maximum gains are
achieved at $\Delta f = 0$~cm. However, for FAST, a focal shift of
about 4.8~cm is needed for the best gain. This small shift is quite
critical for the performance at 3~GHz, not only showing that telescope
gain could be improved by 0.6~dB, but also indicating that the
tracking of a radio source for flux measurements needs good stability
in the focal distance control.

The beam patterns calculated using the official values of $\rho_c=318.5$~m
and $\Delta f=0$~cm are compared with those calculated with the best
values in Fig.~\ref{fig8}. The shapes of the central beam are very
different from $z=0\grad$ up to $z=40\grad$ and get much sharper now,
as listed in Table~\ref{tab1}; the telescope gain is 77.76~dBi at $z = 0\grad$,
corresponding to an aperture efficiency of $67.11\%$. Compared to the
beams calculated by using the official values, the
efficiency is improved by $9.6\%$. Around the main beam at $z = 0\grad$,
the pentagram sidelobes are caused by the pentagon-jointed panels of
the main reflector of the FAST \citep[see Fig.2 in][]{dh13}.

\begin{table*}[!hbt]
\caption{Panel offsets $\Delta D$ (unit:mm) at different parts
  ($R=0\sim150$~m) for the best panel positioning of the mimic 300~m
  paraboloid of FAST}
\begin{center}
\begin{tabular} {lrrrrrrrr}
\hline
\hline
$R$             & 0~m    &  20 m   & 40 m   &  60 m   & 80 m   & 100 m   & 120 m & 150 m \\
\hline
$\rho_c=280$~m   & 0      &  0.4    &  1.0   &  2.3    & 4.7    &  6.2    &  8.3  & 9.2  \\
$\rho_c=300$~m   & $-$3.6 & $-$3.3  & $-$2.3 &  $-$1.4 & 0.9    &  2.6    &  4.8  & 5.7  \\
$\rho_c=310$~m   & $-$5.2 & $-$4.9  & $-$4.0 &  $-$3.0 & $-$0.9 &  0.9    &  3.1  & 4.2  \\
$\rho_c=318.5$~m & $-$6.6 & $-$6.3  & $-$5.4 &  $-$4.4 & $-$2.3 &  $-$0.4 &  1.7  & 2.9    \\
$\rho_c=350$~m   & $-$10.9& $-$10.6 & $-$9.7 &  $-$8.7 & $-$6.9 &  $-$4.6 &$-$2.6 & $-$1.3  \\
\hline
\hline
\end{tabular}
\label{tab2}
\end{center}
\end{table*}


\subsection{Beam optimisation via panel positioning}

\begin{figure}[!hbt]
\begin{center}
\includegraphics[bb=0 228 585 595,width=0.43\textwidth,height=0.26\textwidth,clip]{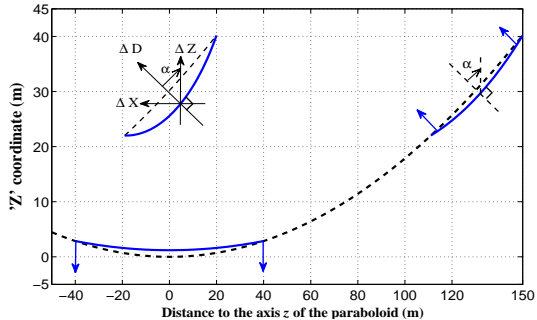}
\caption[]{Illustration of panel offsets in different parts of the expected 300-m paraboloid of the FAST.}
\label{fig9}
\end{center}
\end{figure}

The calculations above and in \citet{dh13} are all based on the
assumption that the three vertexes of each triangular panel sit on a
300-m paraboloid with $f/D$=0.4665 (see Fig~\ref{fig9}). However, by
adjusting panel positions via actutors in the $~2300$ control nodes,
the surface deviation ($rms$ as root-mean-square) of the spherical
panels from the expected paraboloid can be minimized. The panel
``position optimisation''\citep{gj10} is very necessary during the
operation of the FAST.

For each panel, the surface deviation $rms$ should be calculated from
the offset $\Delta D$ in every position in the panel from the expected
paraboloid. The offset $\Delta D$ should be measured perpendicular to
the local surface of the paraboloid (see Fig.~\ref{fig9}). In our FAST
model, considering that every position of each inclined panel has a
very similar inclined angle ($\alpha$) to the vertical $Z$ axis, we can
minimize the surface deviation by searching for one offset in $Z$
directions, $\Delta Z=\Delta D/cos\alpha$, through
%
%
\begin{align*}
min(rms) &=min(\sqrt{\frac{\sum_{i=1}^{n}\Delta D_i^2}{n}})\\
&\approx min(\sqrt{\frac{\sum_{i=1}^{n}\{[(Z^s_i+\Delta Z)-Z^p_i]\cos{\alpha}\}^2}{n}})\tag{1}
\end{align*}
In our calculation for each panel, about n$\approx$200 equally-spaced
points are sampled. $Z^s_i$ and $Z^p_i$ are the '$Z$' coordinates of
the spherical panel and expectd paraboloid at the $i$th sampling
point. This makes our calculation easy, while in practice actutors
have to adjust the positions of the cable net to the expected
paraboloid with a shift of $\Delta D=\Delta Z\cos\alpha$. We derived
the values of $\Delta D$ which make the $rms$ smallest for panels in
various part of the expected paraboloid, as listed in Table.~\ref{tab2}.
The spherical panels should be lower by several mm in the central region,
and lift up by several mm at the edge (see illustration in Fig~\ref{fig9}).

\begin{figure*}[!hbt]
\begin{center}
\includegraphics[bb=2 183 556 640,width=0.325\textwidth,height=0.26\textwidth,clip]{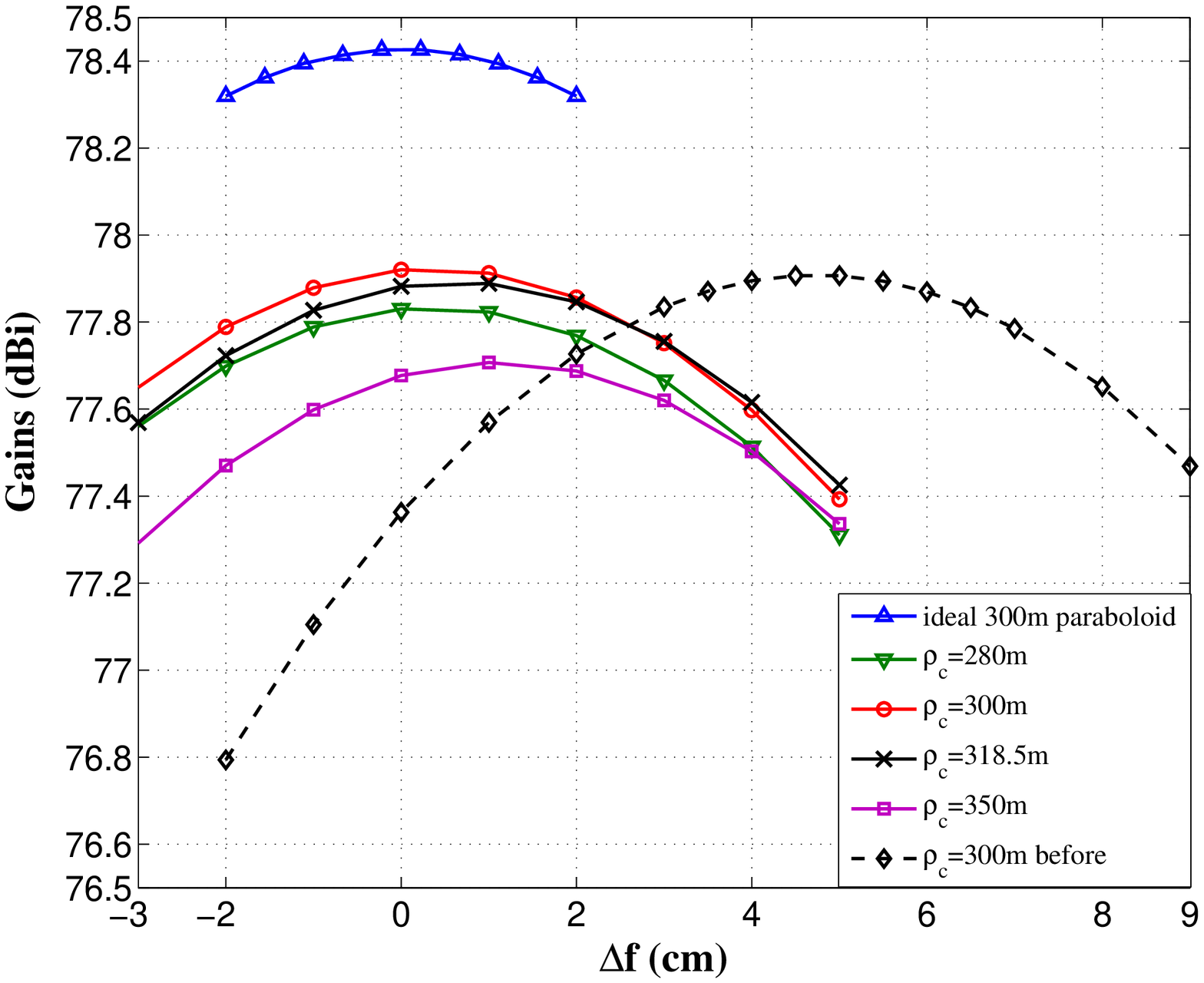}
\includegraphics[bb=31 182 562 644,width=0.315\textwidth,height=0.26\textwidth,clip]{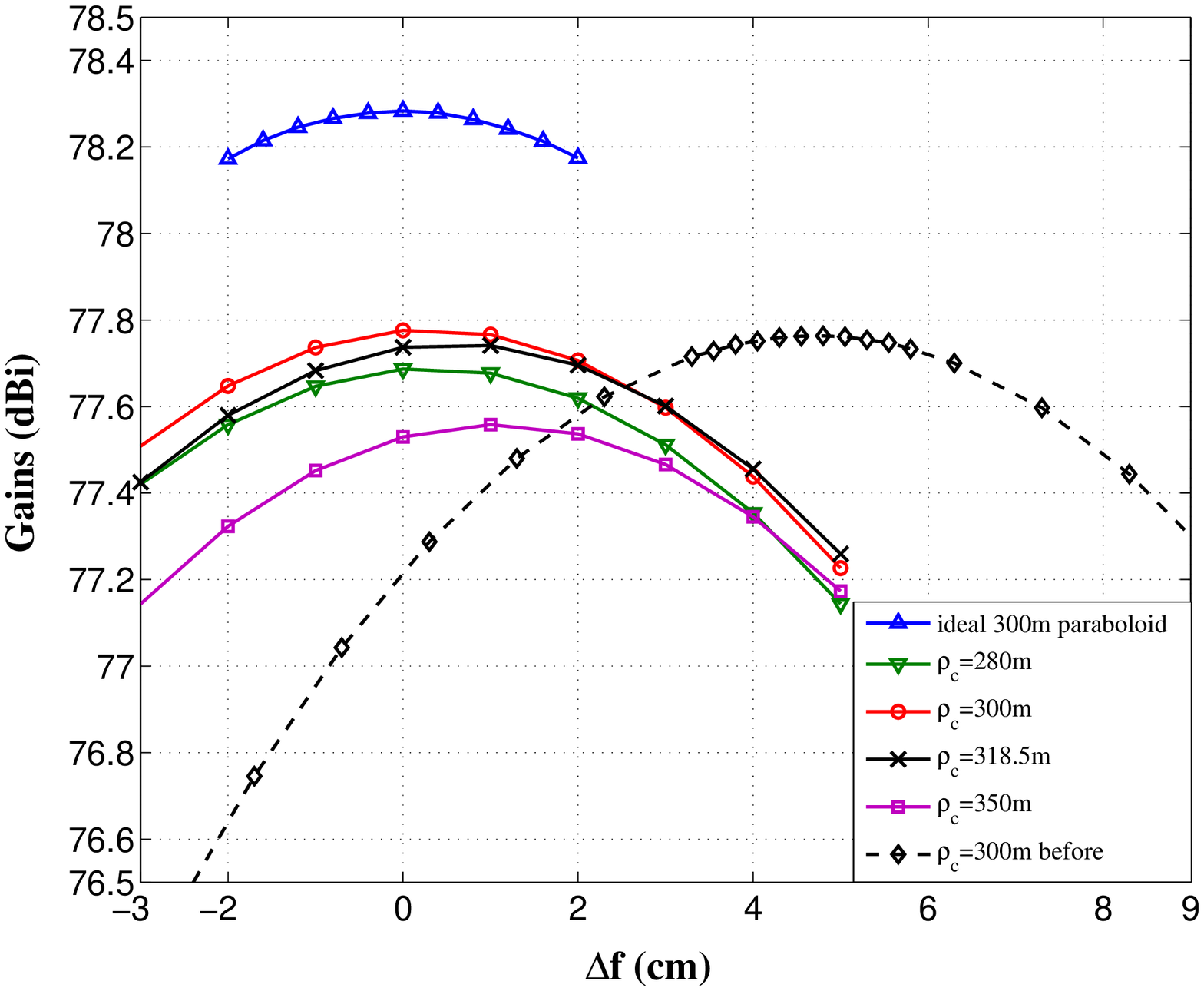}
\includegraphics[bb=29 184 559 643,width=0.315\textwidth,height=0.26\textwidth,clip]{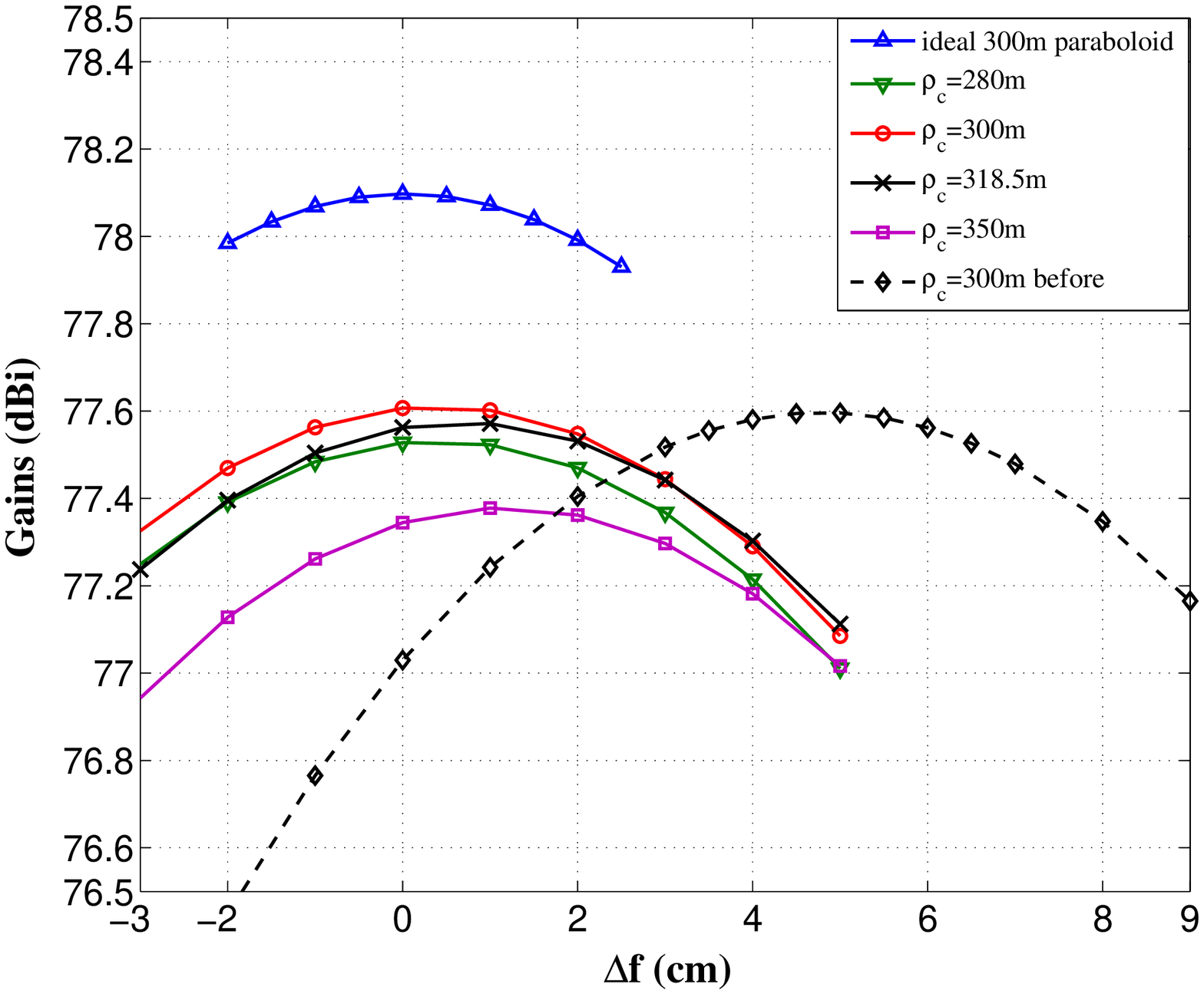}\\[2mm]

\subfigure[$T_{e}$=$-$9.6 dB]{
\includegraphics[bb=6 177 552 642,width=0.3275\textwidth,height=0.28\textwidth,clip]{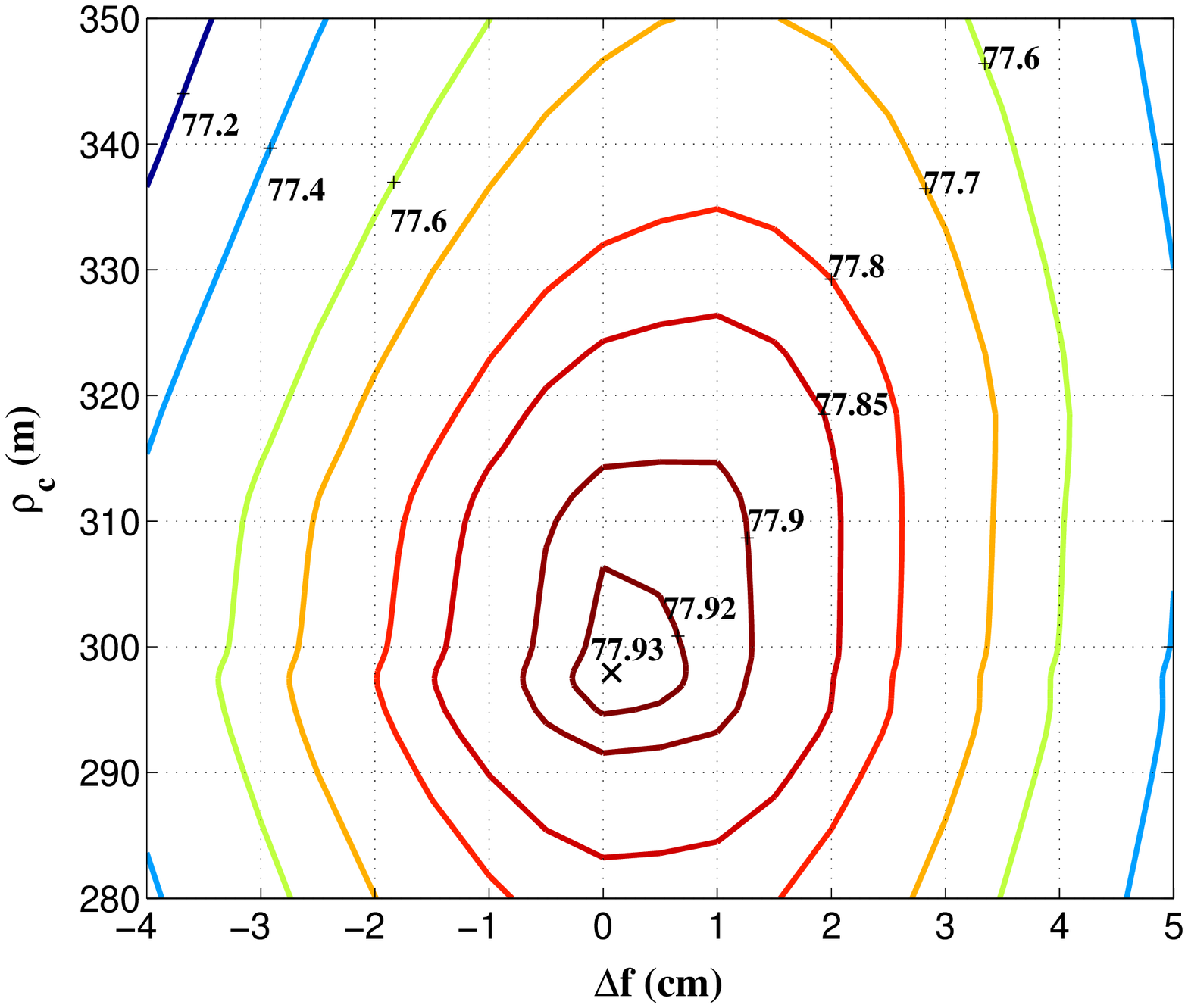}}
\subfigure[$T_{e}$=$-$10.7 dB]{
\includegraphics[bb=43 180 552 643,width=0.3105\textwidth,height=0.28\textwidth,clip]{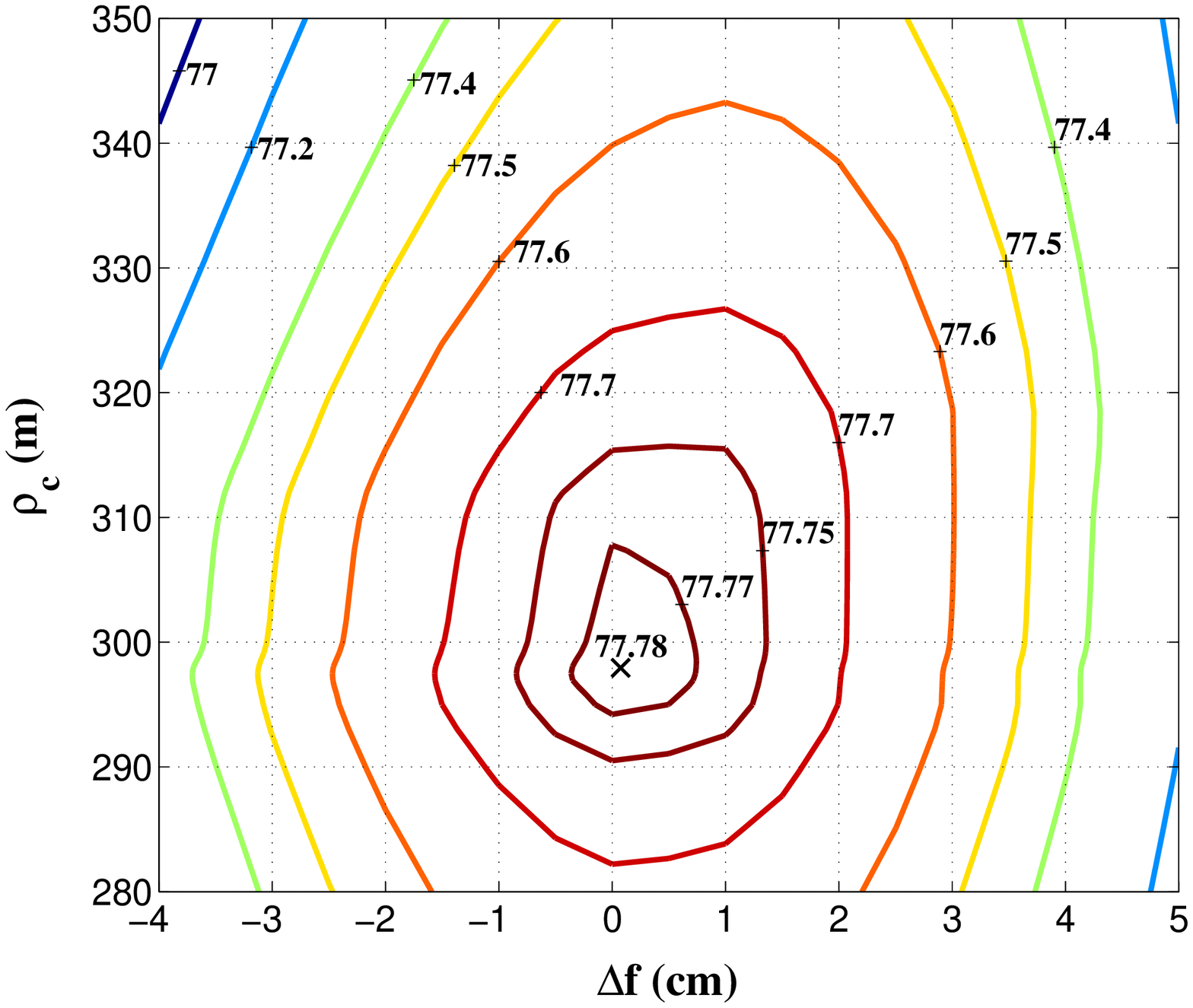}}
\subfigure[$T_{e}$=$-$12 dB]{
\includegraphics[bb=41 182 550 642,width=0.315\textwidth,height=0.28\textwidth,clip]{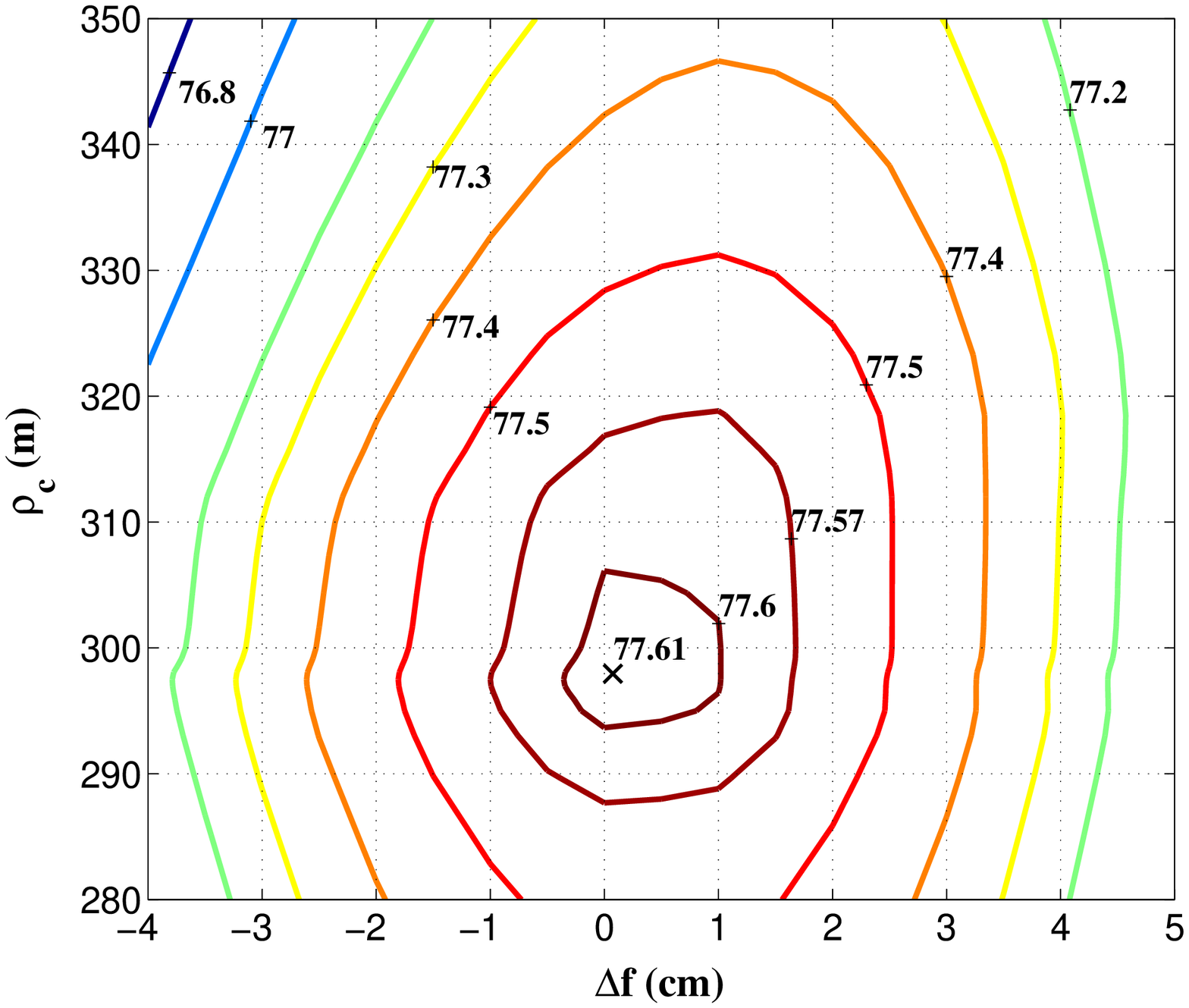}}
  \caption{After the best positioning of panels, gain curves are
    calculated as a function of $\Delta f$ for $\rho_c$=280~m, 300~m,
    318.5~m and 350~m in {\it the upper panels}, and gain contours for
    $\Delta f$ and $\rho_c$ in {\it the lower panels} are calculated for
    three different edge tapers of the feed: $T_e=-9.6$~dB,
    $T_e=-10.7$~dB and $T_e=-12.0$~dB. For comparison, the gain for
    the ideal 300-m paraboloid as a function of the focal shift $\Delta
    f$ is calculated and plotted as blue lines in {\it the upper
      panels}.}
  \label{fig10}
\end{center}
\end{figure*}

Using this new FAST model of adjusted panels, we calculate the beam
patterns and telescope gains at 3~GHz for observations at zenith
angles of $z = 0\grad$, $z = 27\grad$ and $z = 40\grad$. Gain curves
as a function of $\Delta f$ for $\rho_c$=280~m, 300~m, 318.5~m and
350~m are calculated as shown in Fig.~\ref{fig10}. These curves exhibit
almost the same best performance as that in Fig.~\ref{fig7} but the
focal shift needed is very small, in the range of $\Delta f= 0 \sim
1$~cm. The best curvature radius (see Fig.~\ref{fig10}) is also found
to be $\sim$300~m for the maximum gain, very similar to that in
Fig.~\ref{fig7}. The beam pattern is also very similarly optimised as
shown in Fig.~\ref{fig8}. This means that the precision in positioning
of panels to achieve an accuracy of 1~mm is needed for the best performance
for the FAST.

\section{Conclusions}

Our calculations of the beam patterns for the FAST model with a
practical feed show that the best value of the curvature radius of
spherical panels should be $\sim300$~m. The feed should shift
4.8~cm up to get the best gain and best beam shapes. These values
are slightly different from the official settings of the FAST \citep{nan11,gj10},
but not very significant. However, we see that the aperture efficiency
could be improved by $\sim$10$\%$ at 3~GHz and that the beam shapes
become sharper with the optimised panel curvature and feed position,
in addition to the gain increasing by 0.6~dB. We also tried the
best positioning of panels, and found that similar best beam patterns
could be achieved through adjusting panel positions by a few mm. Our
beam calculation results suggest that in future FAST tracking
observations, an excellent stability and accurate positioning
of feed movements to a few mm and of panels to 1~mm or better, have
to be provided so that the observational beams do not obviously vary.

\section*{Acknowledgments} 
The authors are supported by the National Nature Science Foundation of China (10833003),
and thank CST for support.

\end{document}